\documentclass[twocolumn, usenames, dvipsnames]{aastex63}
\usepackage{amsmath,amsthm,amsfonts,amssymb,bm}
\usepackage{physics,commath}

\usepackage{color}
\usepackage{listings}
\usepackage{tikz}
\usepackage{graphicx,comment}
\usepackage{textcomp}
\usepackage{epstopdf}
\usepackage{natbib}
\usepackage{xlmacros}
\usepackage{soul}
\usepackage{comment}

\begin{document}

\title{
Three-Dimensional Reconstruction of Weak-Lensing Mass Maps \\
with a Sparsity Prior. II.
Weighing Triaxial Cluster Halos
}
\author{Shouzhuo Yang}
\email{yangsz@ihep.ac.cn}
\affiliation{Physics and Astronomy Department, Swarthmore College, Swarthmore,
PA 19081}
\affiliation{Institute of High Energy Physics, Chinese Academy of Sciences, Beijing 100049, China}
\author{Xiangchong Li}
\affiliation{Department of Physics, McWilliams Center for Cosmology, Carnegie
Mellon University, Pittsburgh, PA 15213, USA}
\affiliation{Kavli Institute for the Physics and Mathematics of the Universe
(WPI), University of Tokyo, Chiba 277-8583, Japan}
\author{Naoki Yoshida}
\affiliation{Kavli Institute for the Physics and Mathematics of the Universe
(WPI), University of Tokyo, Chiba 277-8583, Japan}
\affiliation{Department of Physics, University of Tokyo, Tokyo 113-0033, Japan}
\affiliation{Institute for Physics of Intelligence, University of Tokyo, Tokyo
113-0033, Japan}

\begin{abstract}
Continuing work presented in \cite{massmap_Li2021}, we performed a series of
tests to our high-resolution three-dimensional mass map reconstruction
algorithm \splinv{}. We test the mass reconstruction accuracy against
realistic mock catalogs generated using shear field produced by triaxial
halos with the inner density profile of $\rho \propto r^{-1}$
and of $\rho \propto r^{-1.5}$.
The galaxy shape noise is modeled
based on the Year-1 Subaru Hyper Suprime-Cam (HSC) Survey. After reviewing
mathematical details of our algorithm and dark matter halo models, we
determine an optimal value of the coefficient of the adaptive LASSO
regression penalty term for single halo reconstruction. We
successfully measure halo masses for massive triaxial halos;
the mass determination accuracy is
5 percent for halos with $M = 10^{14.6}~M_\odot$ at $0.0625\leq z \leq 0.2425$, and 5 percent for those with $10^{14.8}~M_\odot$
at $0.0625\leq z \leq 0.4675$, and 20 percent for $M= 10^{15.0} ~M_\odot$ and
$M=10^{15.2}~M_\odot$ in the redshift range $0.0625\leq z \leq 0.4675$. The redshift estimate accuracy is consistently below $\Delta z /z \leq 0.05$ for
the above halo masses in the range $0.1525\leq z \leq 0.4675$. We also
demonstrate that the orientation of triaxial halos and systematic error in
our halo model do not affect reconstruction result significantly. Finally,
we present results from reconstruction of mass distribution
using shear
catalogs produced by multiple halos, to show \splinv{}'s capability using realistic shear maps from ongoing and future galaxy surveys.
\end{abstract}
\keywords{gravitational lensing: weak — galaxies: clusters: general}

\section{INTRODUCTION}
Galaxy clusters
are the
heaviest gravitationally bound objects in the Universe. The redshift evolution of the
abundance of galaxy clusters is sensitive to the growth rate of cosmic large-scale
structures and the expansion history of the Universe. By reconstrucing the number
of dark matter halos at certain redshift with certain mass and comparing with
halo mass function models (e.g. \citealt{Tinker2010, haloMassFunc_Despli2016}),
one can constrain the underlying cosmological parameters including $\sigma_8$
and $\Omega_m$\,. Cluster cosmology will be one of the main focuses of
upcoming galaxy surveys including \emph{Euclid} and Rubin LSST (e.g., see
\citealt{EuclidLaureijs} and \citealt{LSSTOverviwe2019}).

Gravitational lensing refers to the distortion of light from background
galaxies due to foreground gravitational potentials. Neglecting the $B$-mode in
lensing distortion (which is three orders of magnitudes smaller than $E$-mode,
see \citealt{BmodeLensing_Krause2010}), this effect is usually described by 3
parameters, a spin-0 convergence $\kappa$ and two components of spin-2 shear
$\gamma = \gamma_1 + i\gamma_2$, where convergence changes apparent galaxy
sizes, and shear anisotropically distorts galaxy shapes. By measuring the
coherent anisotropy in galaxy shapes, one can infer the local shear statistically.

Due to the ubiquitousness of this signal, one can measure it from every
detected galaxy at different positions and obtain a map of
shear field \citep{FPFS2022b}. Then the convergence map can be reconstructed using the shear map
\citep{massMap_KS1993}. This convergence map is also known as two-dimensional
(2-D) lensing mass map since it is the integrated foreground mass map along the
line-of-sight (weighted by lensing kernel).

There are extensive studies on $2$-D mass map reconstructed from weak
gravitational lensing shear measurement, which focus on directly detecting
galaxy clusters from weak lensing mass map without modeling the relation
between optical observables and dark matter halo mass
\citep{massMap_HSC1_cluster, massMap_HSC1_cluster2, massMap_HSC3_cluster}.
These studies detect clusters by finding peaks in the reconstructed 2-D mass
map. However, 2-D lensing mass map does not provide redshift and mass
information of galaxy clusters. Therefore, we cannot use a 2-D mass map to
directly study the {\it redshift evolution} of halo mass function.

This paper focuses on detecting and weighing galaxy clusters from
three-dimensional (3-D) mass map reconstructed from weak lensing shear
measurements using the algorithm proposed by \citet{massmap_Li2021}. The
sparsity regularization --- adaptive LASSO \citep{AdaLASSO-Zou2006} ---
utilized by our reconstruction should solve smearing problem of the
reconstructed structures along the line of sight \citep{massMap_HST_Massey2007,
massmap_Hu2002}. We model the 3-D mass map as a sum of basis ``atoms'' in
comoving coordinates as a given 3D density field. The basis ``atoms'' are
constructed with NFW \citep{halo_nfw} or cuspy NFW \citep{halo_JS02} halos,
which differs from other reconstruction schemes as GLIMPSE
\citep{Glimpse3D_Leonard2014} that our basis can accounted for the angular
scale difference at different lens redshifts and is better suited to model
clumpy mass distribution.

By numerically calculating the shear field produced by NFW and cuspy NFW halos
and adding realistic noises from Hyper Supreme Cam (HSC) first-year survey
\citep{Mandelbaum2017_HSCY1}, we attempt to recontruct the underlying halo mass
using ``atoms'' with density profiles described by \citet{halo_OLS03}.

This paper is organized as follows: In Section~\ref{sec:massmap}, we introduce
our algorithm for $3$-D mass map reconstruction. In Section~\ref{sec:onehalo},
we study the cluster detection and cluster mass, redshift estimation from $3$-D
mass map using one-halo simulations with different triaxial profiles. In
Section~\ref{sec:twohalos}, we study the performance of $3$-D mass map
reconstruction using two-halos simulations. In Section~\ref{sec:Sum}, we
summarize and discuss the future application of the method to weak lensing
imaging surveys.

In this paper, we adopt the $\Lambda$CDM cosmology of the Planck 2018
observation of the cosmic microwave background (CMB) with $H_0=67.4
~\rm{km~s^{-1} Mpc^{-1}}$, $\Omega_m=0.315$, $\Omega_\Lambda=0.685$,
$\sigma_8=0.811$, $n_s=0.965$ and $N_{\text{eff}}=2.99$
\citep{cmb-Planck2018-Cosmology}.

\section{$3$D MASS MAP RECONSTRUCTION}
\label{sec:massmap}
In this section, we review the 3-D mass map reconstruction
algorithm introduced in \citet{massmap_Li2021}.

\subsection{Forward modelling}
\label{sec:massmap_forward}

Under the usual Born approximation \citep{born_Petri2017}, the weak lensing shear
field, $\gamma$, observed from background galaxy images is related to the
foreground density contrast field $\delta_\vtheta= \rho_\vtheta
/\bar{\rho_\vtheta}-1$ through a linear transform:
\begin{equation}\label{eq-delta2gamma}
    \gamma_{\vtheta} =
    \sum_\vtheta T_{\vtheta\vtheta'} \delta_{\vtheta'} + \epsilon_\vtheta\,,
\end{equation}
where $\epsilon_\vtheta$ is the error in shear measurement due to the random
galaxy shapes (intrinsic shape noise) and the sky variance (photon noise). Here
$\gamma_\vtheta$, $\delta_\vtheta$ and $\epsilon_\vtheta$ are functions of
$\vtheta\,$, and $T_{\vtheta \vtheta'}$ is a linear mapping operator from
density contrast field to shear field.

In order to reconstruct high-resolution mass maps with hight signal-to-noise
ratio (SNR), we incorporate prior information on the density contrast field
into the reconstruction by modeling the density field as a sum of basis atoms
in a ``dictionary'':
\begin{equation}\label{eq-x2delta}
    \delta_{\vtheta} = \sum_{\vtheta'}\Phi_{\vtheta\vtheta'} x_{\vtheta'},
\end{equation}
where $\Phi_{\vtheta\vtheta'}$ is the matrix transforming from the projection
coefficient vector $x_\vtheta$ to the density contrast field $\delta_\vtheta$.
Note that a dictionary may contain multiple ``frames'', used to contain halos with different density profiles.
The column vectors of $\Phi_{\vtheta\vtheta'}$ are the basis ``atoms'' of the
model dictionary.

We define the forward transforming matrix $A_{\vtheta\vtheta'}=
\sum_{\vtheta''}T_{\vtheta\vtheta''} \Phi_{\vtheta'' \vtheta'}\,.$ With
Equations~\eqref{eq-delta2gamma} and \eqref{eq-x2delta}, and write the
transform from the coefficient vector $x$ to the observed lensing shear field
as
\begin{equation}\label{eq-x2gamma}
    \gamma_{\vtheta}=\sum_{\vtheta'}A_{\vtheta \vtheta'} x_{\vtheta'}
    + \epsilon_{\vtheta}\,.
\end{equation}

To simplify the equations in the following, we use Einstein notation:
\begin{equation}
\begin{split}
  A_{\vtheta \vtheta'} x_{\vtheta'}
  &= \sum_{\vtheta'} A_{\vtheta \vtheta'} x_{\vtheta'}\\
  T_{\vtheta\vtheta''} \Phi_{\vtheta'' \vtheta'}
  &= \sum_{\vtheta''}T_{\vtheta\vtheta''} \Phi_{\vtheta'' \vtheta'}\,.
\end{split}
\end{equation}

\subsection{Sparsity regularization}
\label{sec:massmap_reg}

To obtain a sparse reconstruction of mass map, we use the  $l^1$ norm of the
projection coefficient vector to regularize the modeling. The estimator is
defined as
\begin{equation}
\label{eq:lasso}
\hat{x}_{\vtheta}^{\rm{LASSO}}=\argmin_{x_\vtheta'} \left\{
\frac{1}{2} {}_\Sigma\norm{(\gamma_\vtheta - A_{\vtheta \vtheta'} x_{\vtheta'} )}_2^2+
    \lambda \norm{x_{\vtheta'}}^1_1\right\},
\end{equation}
where $\norm{\bigcdot}_1^1$ and $\norm{\bigcdot}_2^2$ refer to the $l^1$ norm
and $l^2$ norm, respectively, and $\lambda$ is the penalty parameter for the
LASSO estimation. The $l^2$ norm is calculated with the inverse of covariance
matrix of the shape noise in the shear measurement:
$\mathcal{C}_{\vtheta\vtheta'}=\langle
\epsilon_{\vtheta}\epsilon_{\vtheta'}\rangle\,$.
As we do not smooth the observed shear map across pixels,
$\mathcal{C}_{\vtheta\vtheta'}$ and its inverse are approximately diagonal.

\begin{figure}
\centering
\includegraphics[width=0.48\textwidth]{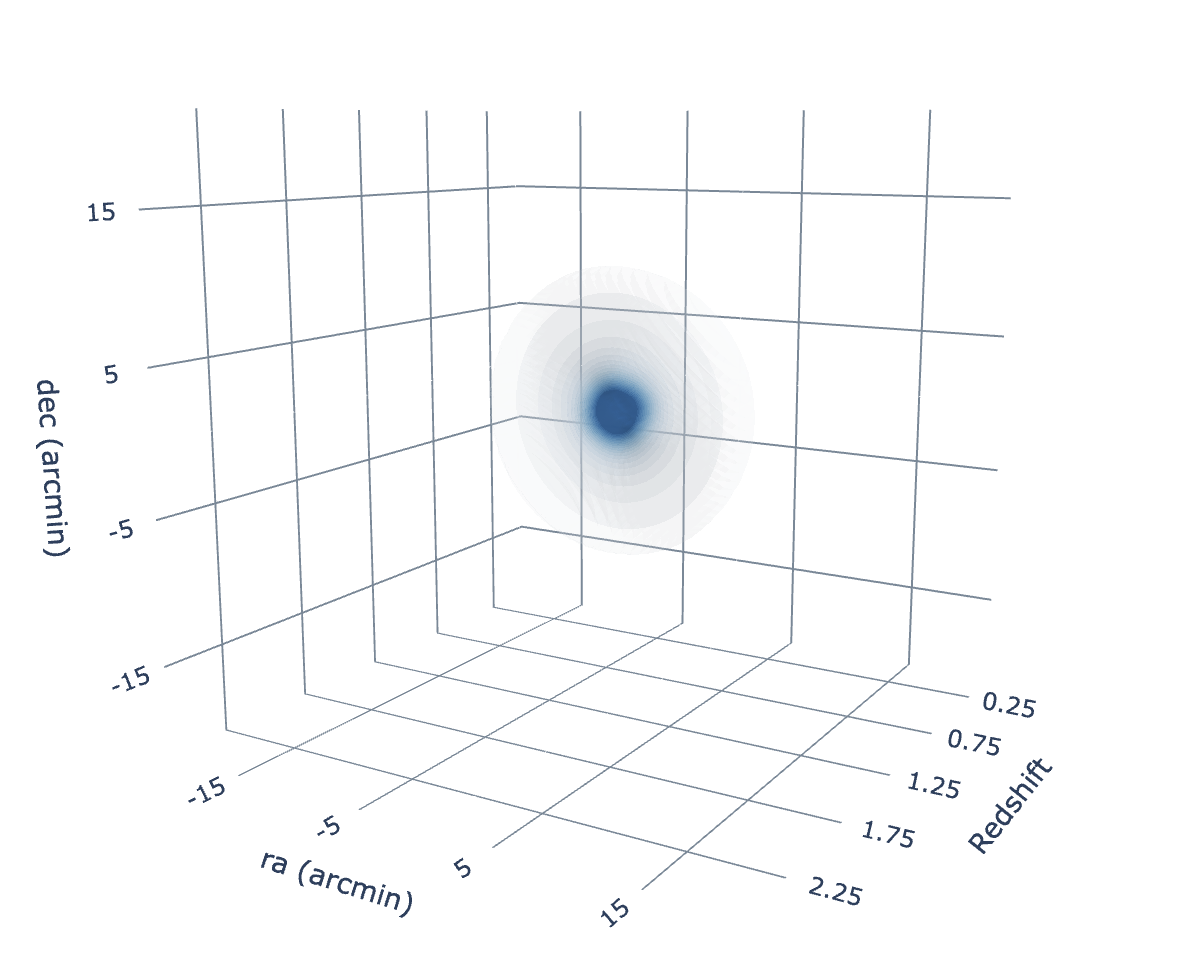}
\caption{
    A sample reconstruction illustration of an isotropic halo at $z=1.25$ with
    mass $M = 10^{14.6}M_{\odot}$.
    }
    \label{fig:sample_1halo}
\end{figure}

\begin{figure}
\centering
\includegraphics[width=0.48\textwidth]{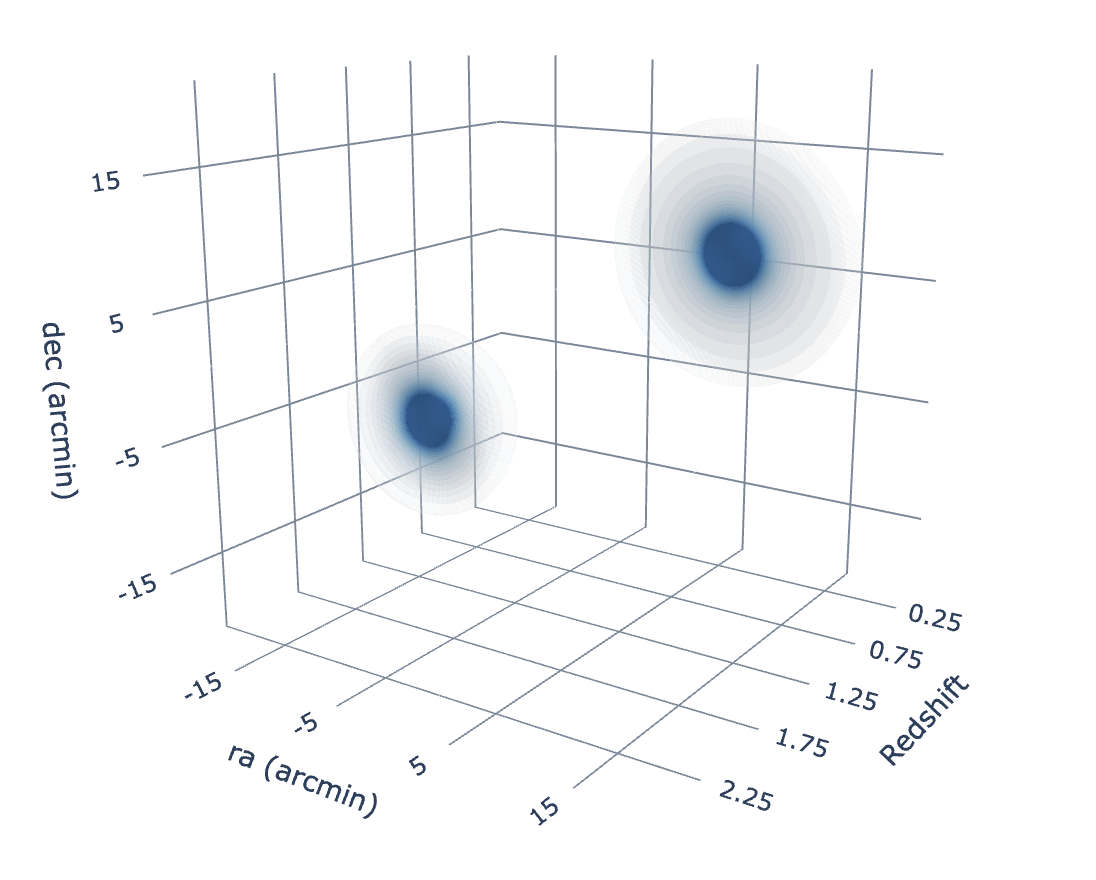}
\caption{
    A sample reconstruction of underlying $\kappa$ map of a 2-halo simulation
    using two isotropic halos at $z=2.25$ with $M = 10^{14.8}M_{\odot}$ and
    $z=1.25$ with $M = 10^{14.6}M_{\odot}$\,.
    }
    \label{fig:sample_2halo}
\end{figure}

The LASSO algorithm searches and selects the parameters that are relevant to
the measurements, and simultaneously estimates the values of the selected
parameters. It has been shown by \citet{AdaLASSO-Zou2006} that when the column
vectors of the forward transform matrix $\mathbf{A'}$ are highly correlated,
the algorithm cannot select the relevant atoms from the dictionary
consistently. In addition, the estimated parameters are often biased owing to
the shrinkage in the LASSO regression. We note that, for the density map
reconstruction problem here, the column vectors are highly correlated even in
the absence of photo-$z$ uncertainties since the lensing kernels for lenses at
different redshifts overlap significantly \citep{massmap_Li2021}. Therefore,
the LASSO algorithm cannot precisely determine the consistent mass distribution
in redshift, and the reconstructed map suffers from smearing in the line of
sight direction even in the absence of noises.

To overcome the problems, \citep{massmap_Li2021} adopts the adaptive LASSO
algorithm proposed in \citet{AdaLASSO-Zou2006} proposes, which uses adaptive
weights to penalize different projection coefficients in the $l^1$ penalty. The
adaptive LASSO algorithm performs a two-step process. In the first step, the
standard (nonadaptive) LASSO is used to estimate the parameters. We denote
the preliminary estimation as $\hat{x'}_{\rm{LASSO}}$. In the second step, the
preliminary estimate is used to calculate the non-negative weight vector for
penalization as
\begin{equation}
\hat{w}_\vtheta= \frac{1}{\abs{\hat{x'}_{\rm{LASSO}}}_{\vtheta}^\tau},
\end{equation}
where we set the hyperparameter $\tau$ to $2\,$ \citep{massmap_Li2021}. The
adaptive LASSO estimator is then given by
\begin{equation}\label{eq-lossFun}
\hat{x}_{\vtheta}=\argmin_{x_\vtheta'} \left\{
\frac{1}{2} {}_\Sigma\norm{(\gamma_\vtheta - A_{\vtheta \vtheta'} x_{\vtheta'} )}_2^2+
\lambda_\text{ada} \norm{\hat{w}_{\vtheta'} \circ x_{\vtheta'}}^1_1\right\},
\end{equation}
where ``$\circ$'' refers to the element-wise product. $\lambda_{\rm{ada}}$ is
the penalty parameter for the adaptive LASSO, which does not need to be the
same as the penalty parameter for the preliminary LASSO estimation $\lambda$.
The adaptive weights enhance the shrinkage in the soft thresholding for the
coefficients with smaller amplitudes, whereas the weights suppress the
shrinkage for the coefficients with larger amplitudes.

We show examples of reconstructed mass map on halo simulations with one input
halo and two input halos in Figs.~\ref{fig:sample_1halo} and
\ref{fig:sample_2halo}, repectively. The details for these two cases will be
discussed in Sections~\ref{sec:onehalo} and \ref{sec:twohalos}, repectively.

\begin{figure}[!ht]
\centering
\includegraphics[width=0.5\textwidth]{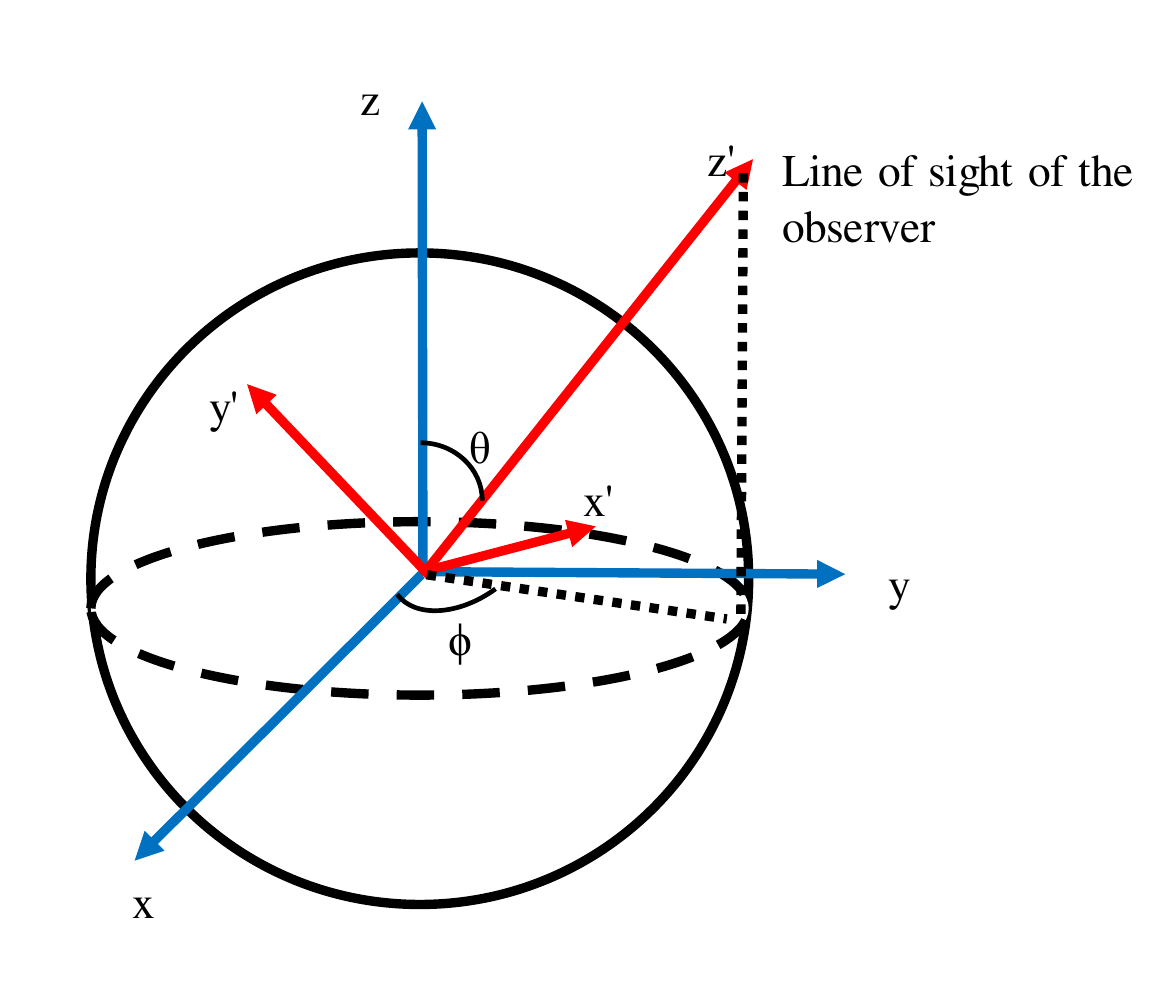}
\caption{
    Orientations of the coordinate systems. The Cartesian axes $(x,y,z)$
    represent the halo principal coordinate system, while the axes $(x', y',
    z')$ stand for the observer’s coordinate system, with the $z'$ -axis
    aligned with the line-of-sight direction. The $x'$-axis lies in the $x$-$y$
    plane. The angle $\theta$, $\phi$ represents the polar angle of the
    line-of-sight direction in the halo's coordinate system. Adapted from
    \cite{halo_OLS03} with permission.
    }
    \label{fig:halo_illustration}
\end{figure}

\section{ONE HALO}
\label{sec:onehalo}

In this section we test the mass map reconstruction on one-halo simulations of
general triaxial halos. In Section~\ref{sec:oneHalo_trihalo}, we present the
density profile and lensing effects of triaxial halos; in
Section~\ref{sec:oneHalo_sim}, we describe the triaxial halo simulations used
to test our mass mapping algorithm; in Section~\ref{sec:oneHalo_noiseless} and
Section~\ref{sec:oneHalo_noisy}, we show the results on noiseless and noisy
simulations, respectively.

\subsection{Triaxial halo Simulation}
\label{sec:oneHalo_trihalo}

\subsubsection{Halo profile}
Following the definition of \cite{halo_JS02}, we adopt a halo density profile
of
\begin{equation}
    \label{eq:trixal_rho_r}
    \rho(R) = \frac{\delta_{\text{ce}} \rho_{\text{crit}}(z)}{(R/R_0)^{\alpha}
    (1+R/R_0)^{3-\alpha}},
\end{equation}
where
\begin{equation}
    R^2\equiv c^2 \left(\frac{x^2}{a^2} + \frac{y^2}{b^2} + \frac{z^2}{c^2}\right)
    \quad \quad (a\leq b \leq c) \,. \label{halo density}
\end{equation}

Here, $\delta_{\mathrm{ce}}$ is the concentration parameter of the halo.
$\rho_{\mathrm{crit}}(z)$ is the critical density of the universe at redshift
$z$. We denote the scale radius of the triaxial halo as $R_0$, and $a,b$ and $c$ are scaling
factors that describes the shape of the halo. We set $b=c=1$ in the rest of the
analysis in this paper, for simplicity and normalization.

From equation~(\ref{eq:trixal_rho_r}), we see that the case of an isotropic
halo model with $\alpha = 1$ reproduces the NFW halo profile. Various
literature uphold different values of $\alpha$ ranging between $\alpha=1$ and
$\alpha=1.5$ (see e.g., \citealt{halo_nfw}, \citealt{halo_Moore_1999},
\citealt{halo_OLS03}). Therefore, we perform our subsequent analyses both for halos with
$\alpha=1$ and $\alpha=1.5$.

\citet{halo_JS02} also define a length scale $R_e$ such that
$R_e/r_{\text{vir}} = 0.45$\, ($r_{\text{vir}}$ is the virial radius) and
$\frac{R_e}{R_0} = c_e$ is the concentration parameter. The average density
within an ellipsoid of $R_e$ is
\begin{equation}
    \Delta_e  = 5\Delta_{\text{vir}} (\frac{c^2}{ab})^{0.75},
\end{equation}
where, from \citet{halo_Oguri_2001}, we have
\begin{equation}
    \Delta_{\text{vir}} = \frac{3M_{\text{vir}}}{4\pi r_{\text{vir}}
    \rho _{\text{crit}}}
    = 18\pi^2 (1 + 0.4093\, \omega_{\text{vir}}^{0.9052})\,.
\end{equation}
Assuming that $\Omega_{m} + \Omega_{\Lambda} = 1$\,, $\omega_{\text{vir}} =
1/\Omega_{\text{vir}} - 1$\,, and the density parameter $\Omega_{\text{vir}}$
at the redshift of virialization $z_{\text{vir}}$ is \citep{Oguri_2001}
\begin{equation}
\Omega_{\text{vir}} = \frac{\Omega_{m} (1+z_{\text{vir}})^3}{\Omega_{m} (1+z_{\text{vir}})^3
    + (1-\Omega_{m} - \Omega_{\Lambda})(1 + z_{\text{vir}}^2) + \Omega_{\Lambda}}\,.
\end{equation}

In this paper, we take the empirical relation between the concentration
parameter ($c \equiv \frac{r_{\mathrm{vir}}}{R_0}$), the viral mass
($M_{\mathrm{vir}}$) and redshift ($z$) of the halo:
\begin{equation}
    c = A \left(\frac{M_{\mathrm{vir}}}{10^{13}M_{\odot}}\right)^B \left(\frac{1.47}{1+z}\right)^C
\end{equation}
where $A = 6.02$, $B=-0.12$, $C = 0.16$, and $M_{\odot}$ is the solar
mass\,\citep{Child2018}.

\begin{figure*}[!ht]
\centering
\includegraphics[width=0.95\textwidth]{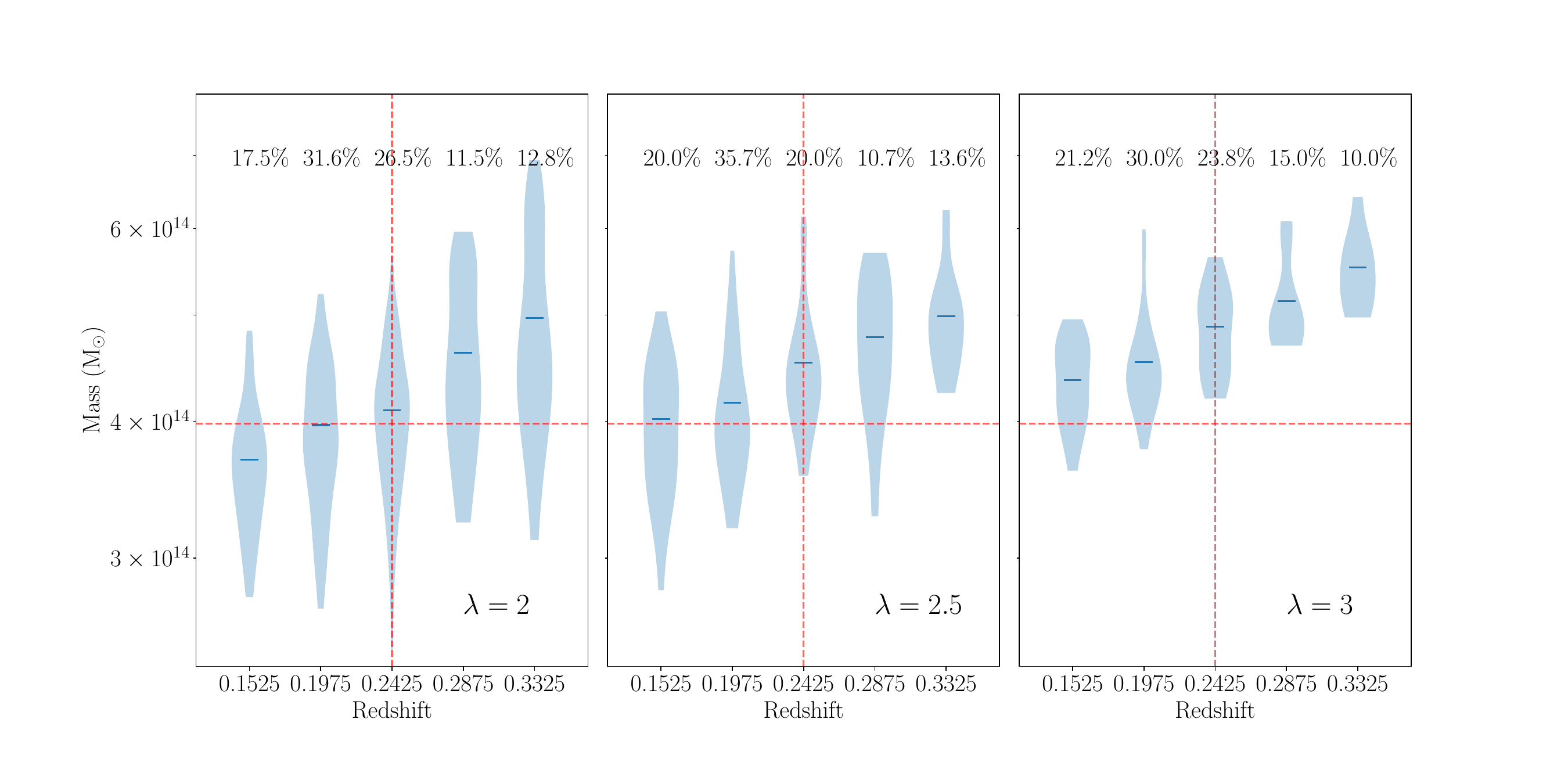}
\caption{
    Detected halo number density as a function of measured redshift and
    measured mass with different sparsity parameters. The shaded blue area
    shows the number density as a function of measured mass located at each
    redshift. The input halo has mass $10^{14.6}~M_{\odot}$. The number above
    each probability distribution shows the fraction of halos detected at the
    corresponding redshift. The detection rates are $47\%$, $28\%$, and $16\%$
    from the left to the right panel, respectively.
    }
    \label{ViolinPlot14.6}
\end{figure*}

\subsubsection{Convergence and Shear of Triaxial Halos}

To describe rotations of a traxial halo, we introduce two coordinate systems:
The $(x, y, z)$ system is the dark matter halo's system, with its origin at
the halo's center and $z$ axis lies along the major principle axis of the halo.
The $(x', y', z')$ coordinate represent the observer's coordinate system, with
its origin also set at the halo's center. We define $(\theta, \phi)$ to be the
polar coordinates of the line-of-sight direction of the observer with the
halo's long axis as the $z$-axis. Just like \cite{halo_OLS03}, we note that the
$x'$-axis lies in the $x$-$y$ plane. We also define $\zeta$ such that
\begin{equation}
    \zeta = \frac{c^2}{b^2} x'^2 + \frac{c^2}{a^2} y'^2,
\end{equation}
where the primed coordinate $(x', y')$ represent normalized observer's
coordinate. Fig. \ref{fig:halo_illustration} is an illustrative plot that shows
the relationship between the coordinates. Then we get an expression for the
lensing convergence $\kappa$ as:
\begin{equation}\begin{aligned}
\label{eq:tiaxial_kappa}
    \kappa=\frac{R_{0}}{\Sigma_{\text {crit }}}
    \int_{-\infty}^{\infty} \rho(R) d z^{\prime}
    &=\frac{R_{0}}{\Sigma_{\text {crit }}} \int_{-\infty}^{\infty}
    \frac{\rho\left(\sqrt{z_{*}^{\prime 2}
    +\zeta^{2}}\right)}{\sqrt{f}} d z_{*}^{\prime} \\
    & \equiv \frac{b_{\mathrm{TNFW}}}{2} f_{\mathrm{TNFW}}(\zeta)\,
\end{aligned}\end{equation}
where
\begin{equation}
    b_{\mathrm{TNFW}}\equiv \frac{1}{f}
    \frac{4\delta_{\mathrm{ce}}\rho_{\mathrm{crit}(z)} R_0}{\Sigma_{\mathrm{crit}}},
\end{equation}
with $f = \sin^2\theta (\frac{c^2}{a^2} \cos^2\phi + \frac{c^2}{b^2}\sin^2\phi)
+ \cos^2\theta $, and
\begin{equation}
    f_{\mathrm{TNFW}} \equiv
    \int^\infty_{0}
    \frac{1}{(\sqrt{r^2 + z^2})^\alpha(1+\sqrt{r^2+z^2})^{3-\alpha}} dz \,.
\end{equation}
The subscripts, ``TNFW'', corresponds to ``Triaxial NFW''. $\Sigma_{\text
{crit}}$ is the lensing critical surface mass density defined as
$$
\Sigma_{\text {crit }} \equiv \frac{c^2 D_{\mathrm{OS}}}{4 \pi G
D_{\mathrm{OL}} D_{\mathrm{LS}}},
$$
where $D_{\mathrm{OL}}, D_{\mathrm{OS}}$,
and $D_{\mathrm{LS}}$ are the angular diameter distances from the observer to
the lens plane, from the observer to the source plane, and from the lens plane
to the source plane, respectively.

Once we have an expression for $\kappa$, we may follow \citet{halo_keeton2001a}
to calculate the shear field. Note that although in $\alpha=1$ case an
analytical solution can be yield, an analytical solution does not exist for a
density profile with $\alpha=1.5$\,. Differing from equation~(4) in
\citet{massmap_Li2021}, we did not adopt truncation at the viral radius to
facility numerical computation of shear fields.

\subsection{Simulation setup}
\label{sec:oneHalo_sim}

\begin{figure*}[!ht]
\centering
\includegraphics[width=0.95\textwidth]{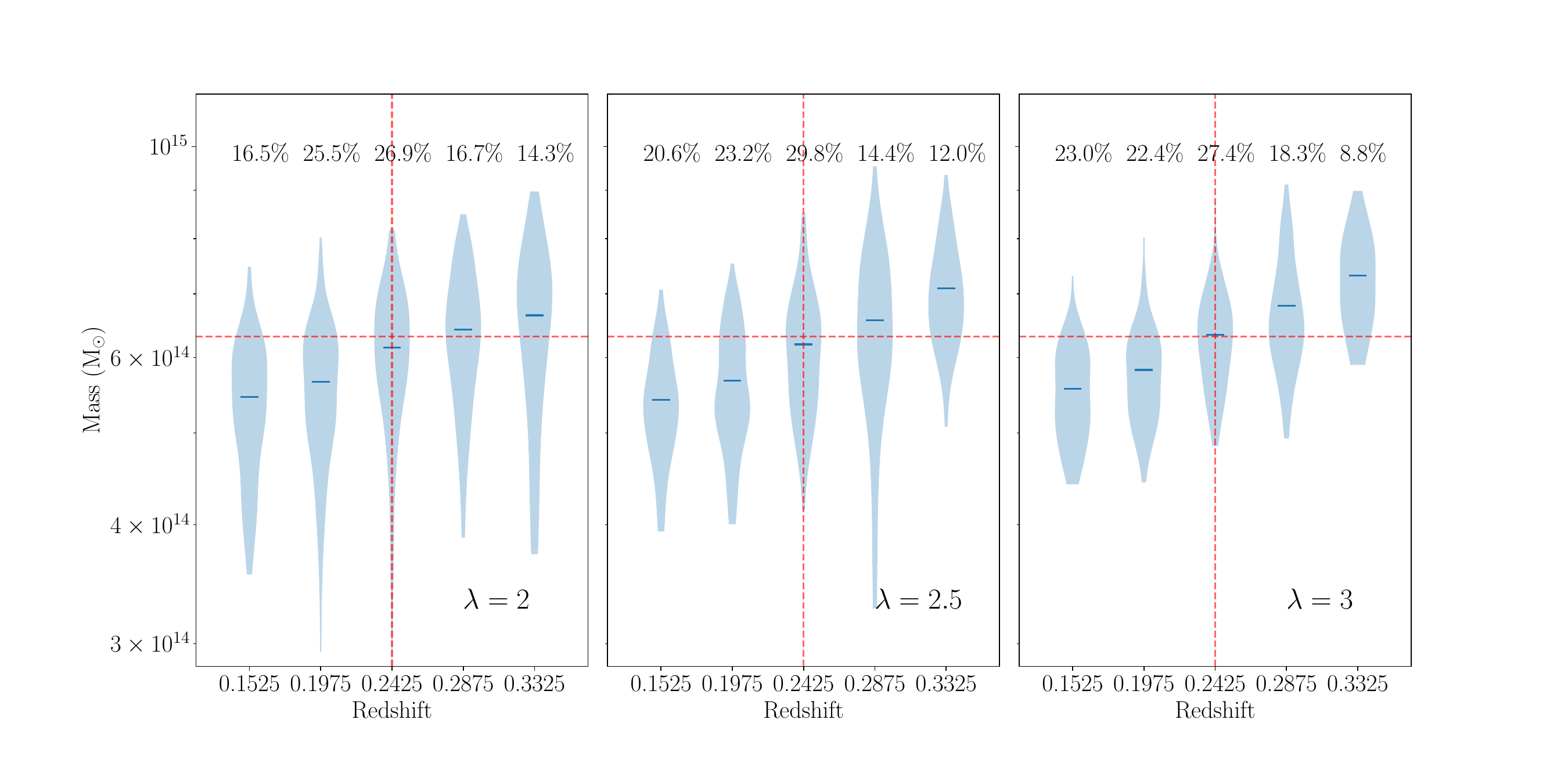}
\caption{
    Similar to Fig.~\ref{ViolinPlot14.6}, but for halos with mass $10^{14.8}
    M_{\odot}$\,. The detection rates are $82\%, 77\%,$ and $68\%$ from the
    left to the right panel, respectively.
    }
    \label{ViolinPlot14.8}
\end{figure*}
\begin{figure*}[!ht]
\begin{center}
    \includegraphics[width=0.9\textwidth]{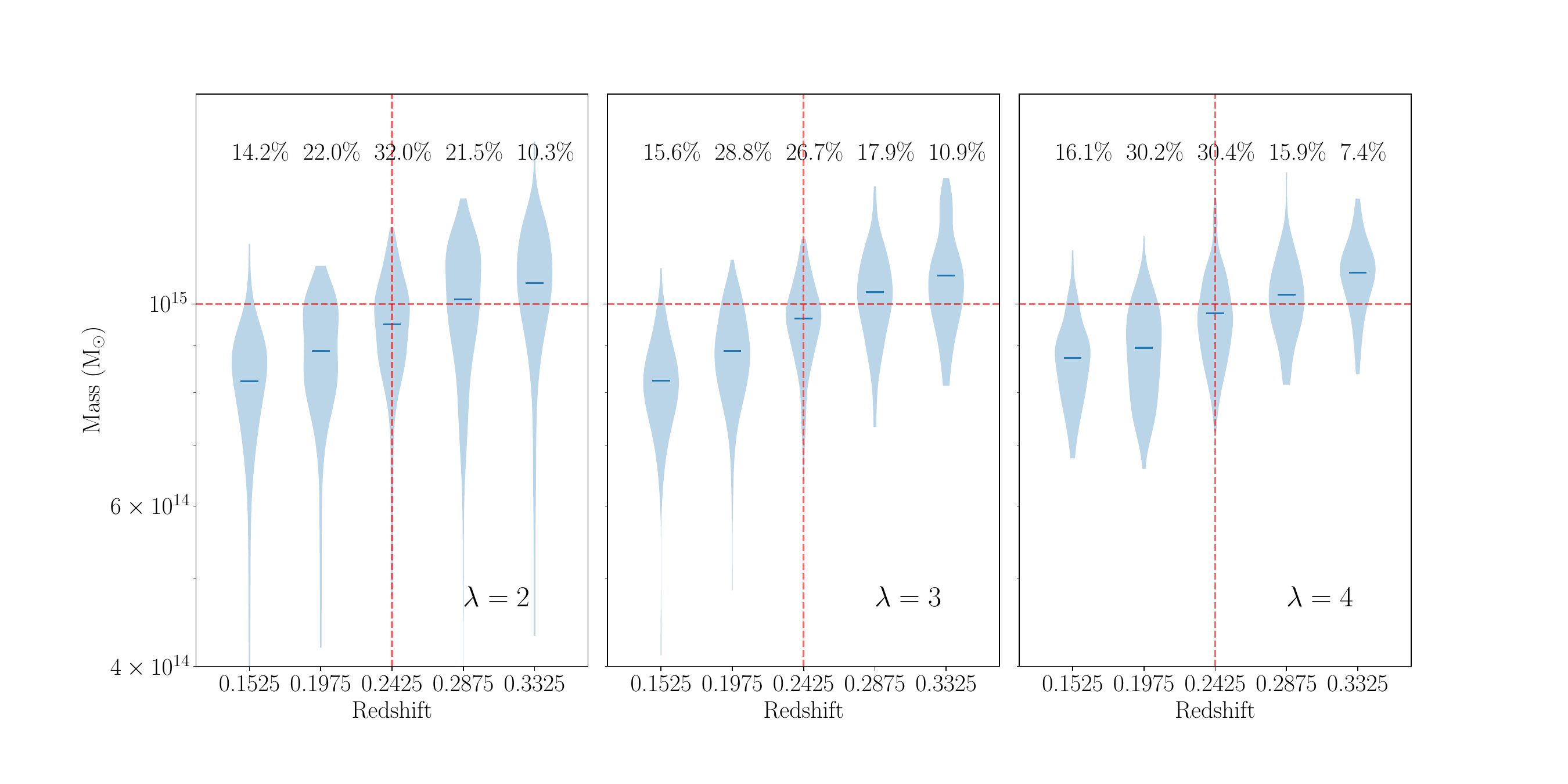}
\end{center}
\caption{
    Similar to Fig.~\ref{ViolinPlot14.6}, but for halos with mass $10^{15.0}
    M_{\odot}$\,. The detection rates are $89\%, 95\%,$ and $95\%$ from the
    left to the right panel, respectively.
    }
    \label{ViolinPlot15.0}
\end{figure*}

In this subsection, we introduce our simulations used to test the mass map
reconstruction algorithm and quantify biases in the halo mass and redshift
estimations from the reconstructed mass map..

\subsubsection{Lensing Shear}
\label{sec:numerical_dictionary}

We use halos with different triaxial shapes to produce shear fields for
simulation. This is represented with a wide range of ellipticity, defined as
$\frac{a}{c}$ (see equation~\ref{halo density}). However, to reduce
dimensionality and computational time during the mass map reconstruction, the
dictionaries are prepared with isotropic halos with $a\kern-0.28em
=\kern-0.28em b \kern-0.28em =\kern-0.28em c \kern-0.28em =\kern-0.28em 1$
described in the previous section. The shear field are measured at redshift of
$z\kern-0.28em =\kern-0.28em 0.05, 0.36, 0.47, 0.56, 0.69, 0.80, 0.91, 1.03,
1.22, 1.50,$ and $2.50$.

To consider variation of halo profiles as found in recent high-resolution
$N$-body simulations \citep{halo_Navarro_2010}, we include the ability to
simulate shear field produced by triaxial halos \citep{halo_JS02} with both (i)
the NFW radial profile with $\alpha=1$ in equation~\eqref{eq:trixal_rho_r}
\citep{halo_nfw} and (ii) the cuspy NFW radial profile with $\alpha=1.5$
\citep{halo_Jing_2000}.

\subsubsection{Observational Noise}
\label{sec:onHalo_noi}

We account for statistical uncertainties in shape estimation from galaxy
intrinsic shape noise and measurement error due to image noise, calculated
using the first-year shear catalog of the HSC first-year data
\citep{Mandelbaum2017_HSCY1}. Specifically, we utilized the formulation of
\cite{shirasaki2019}, where we have
\begin{equation*}
    \epsilon^{\mathrm{int}}=\left(\frac{e_{\mathrm{rms}}}
        {\sqrt{e_{\mathrm{rms}}^2
    + \sigma_{\mathrm{e}}^2}}\right) e^{\mathrm{ran}},
    \quad \epsilon^{\mathrm{mea}}=N_1 + \mathrm{i}\, N_2.
\end{equation*}
In the above expression $\epsilon^{\mathrm{int}}$ represents the per-component
intrinsic shape error, and $\epsilon^\mathrm{mea}$ represents the per-component
shape measurement error. Also, $e^{\mathrm{ran}}=e^{\mathrm{obs}} e^{i \phi}$,
where $\epsilon^{\mathrm{obs}}$ is the distortion of some individual galaxy and
$e^{i\phi}$ serves to rotate the observed shape by some random angle $\phi$.
$N_1$ and $N_2$ are random numbers drawn from a Gaussian centered at $0$ with a
standard deviation of $\sigma_{\mathrm{e}}$. $\epsilon_{\mathrm{rms}}$ is the
root-mean-square of the intrinsic galaxy shape for each shape component.
$\sigma_e$ is the standard deviation of the shape measurement error due to
image noise for each shape component. Note, $\epsilon_{\mathrm{rms}}$ and
$\sigma_{\mathrm{e}}$ are estimated from image simulations at single galaxy
level using realistic galaxy image simulations \citep{HSC1_GREAT3Sim}.

We assume that the multiplicative and additive biases in the shear catalog is
fully corrected in this paper; therefore, we have an expression for the
observed shear:
\begin{equation}
    \boldsymbol{\gamma}^{\mathrm{obs}}=
    \frac{\epsilon}{2 \mathcal{R}}\,,
\end{equation}
where $\mathcal{R} = 1-\left\langle \epsilon_{\mathrm{rms}}^2\right\rangle$. We
can then substitute
$$\epsilon = \epsilon^{\mathrm{ran}} + \epsilon^{\mathrm{mea}}$$
to get a mock shear field.

For realistic noisy tests, we adopt realistic HSC-like galaxy number density
($\sim$20~arcmin$^{-2}$) \citep{HSC1-catalog, HSC3_catalog} when producing
shear fields in order to test the performance of our algorithm with noisy
setup. However, for the noiseless tests in section~\ref{sec:oneHalo_noiseless},
we adopt an extremely high galaxy number density (2000~arcmin$^{-2}$) to
suppress the random noise in the sub-pixel galaxy distribution.

\subsection{Results: Noiseless Case}
\label{sec:oneHalo_noiseless}

\begin{figure*}
\centering
\includegraphics[width=0.9\textwidth]{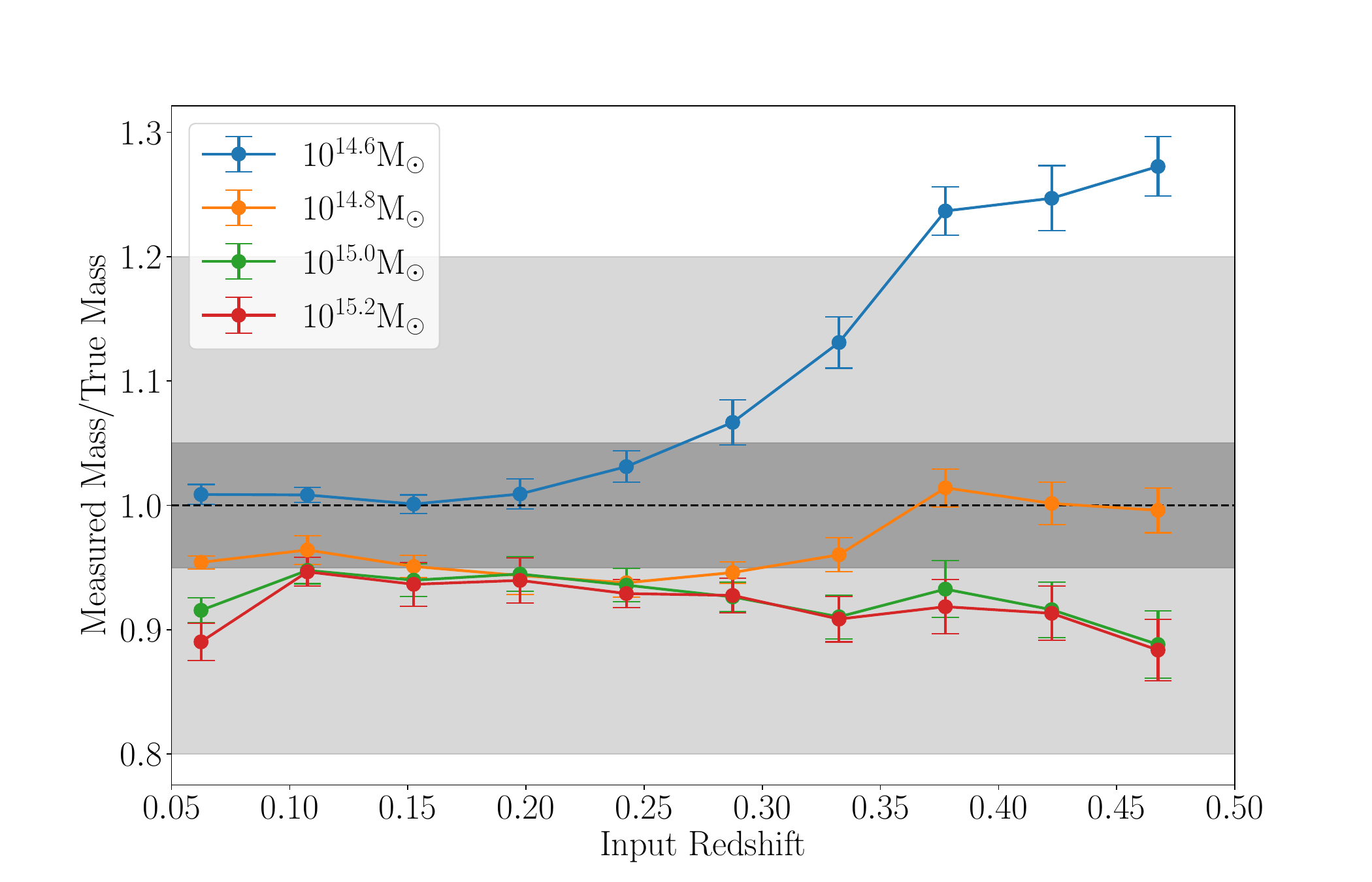}
\caption{
    NFW halo ($\alpha=1$) mass bias for $M=10^{14.6}, 10^{14.8}, 10^{15.0},$
    and $ 10^{15.2}~M_{\odot}$. The mass maps are reconstructed with
    $\lambda=2$. The darker grey area indicate a $5\%$ bias and the lighter
    grey area indicate a $20\%$ relative mass bias. The error bar indicate the
    standard deviation of reconstructed mass with respect to the ellipticity
    $(a/c)$ over the range $[0.5,1]$.
    }
\label{NoisyMassBiasNFWlambda2}
\end{figure*}

In this section, we discuss the mass map reconstruction results for triaxial
NFW ($\alpha = 1$) and cuspy NFW ($\alpha = 1.5$) halos noiseless simulations.
An example of the reconstructed 3-D map is shown in
Fig.~\ref{fig:sample_1halo}.

\begin{enumerate}
    \item Indicating the type(s) of halo(s) from which a dictionary desires to
        be built. This includes the density profile of the halo (in this work,
        we only use NFW or the cuspy NFW halo but there are more
        possibilities), the masses of the halos and either the concentration
        parameter or the scale radius of the halos. Note that a set of
        dictionary may contain a mixture halo models.
    \item  To generate noiseless $\kappa$ field, we prepared a set of function
        to calculate halo's $\kappa$ field from its density profile and other
        properties. We sample equation~\eqref{eq:tiaxial_kappa} with $500$
        points per square arcmin and pixelize the map. Note, when creating the
        $\kappa$ map, we do not smooth to introduce correlation between pixels
    \item Use Kaiser-Squire \citep{massMap_KS1993} transformation to acquire
        the noiseless underlying shear field produced by the halo specified by
        the above parameters.
\end{enumerate}

We first investigated the relative mass bias (defined to be the difference
between true mass and the reconstructed mass over true mass) in \splinv{}'s
estimation from noiseless shear field.  We performed in total of 100
reconstructions, for halos having 10 redshift values from $z=0.0625$ to
$z=0.4675$ and 10 ellipticity values from $\frac{a}{c}=1$ to $\frac{a}{c}=0.5$.
We chose $\lambda = 2$ because this value is our fiducial value for later noisy
reconstruction and specific value of $\lambda$ does not affect reconstruction
result significantly in this noiseless reconstruction. We repeated the above
procedure for three masses: $M_{\mathrm{vir}} = 10^{14.6}~M_\odot$,
$10^{14.8}~M_\odot$, and $10^{15.0}~M_\odot$\,. We put each halo in the center
of in a $48 \times 48$ pixelized grid covering $98 \,\text{arcmin} \times 98
\,\text{arcmin}$ of sky area. Our result shows that near-isotropic halo
reconstruction gives exquisite mass estimation: for halos with $0.8\leq
\frac{a}{c} \leq 1$, mass bias is consistently around or below percent-level
across $z=0.0625$ to $z=0.4675$. We conclude that the noiseless reconstruction
at a lower redshift ($z\leq 0.2425$) only has $~5\%$ of mass bias even when the
halo is highly anisotropic, while the highest mass bias at $z=0.3325$ is less
than $~10\%$. There are some instances where \splinv{} overestimates the
redshift of halos with a small $\frac{a}{c}$ in high redshifts. We think this
is due to the fact that a smaller $\frac{a}{c}$ values correspond to a smaller
halo as it appears along the line-of-sight direction, and therefore the
\splinv{} will tend to approximate the field with halo with a smaller radius,
corresponding to a higher redshift.

\subsection{Results: Noisy Case}
\label{sec:oneHalo_noisy}

\begin{figure*}
\centering
\includegraphics[width=0.9\textwidth]{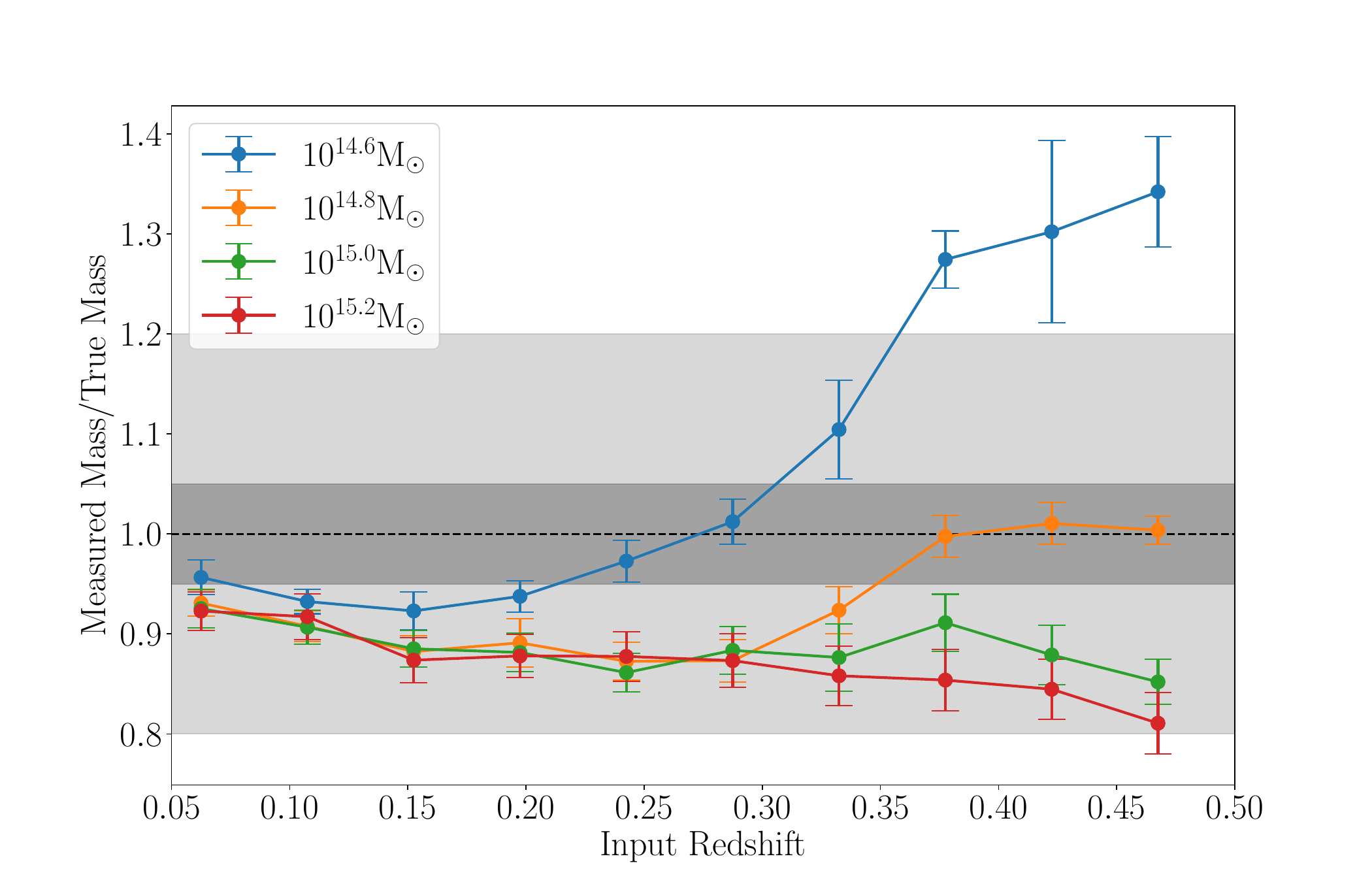}
\caption{
    Cuspy NFW halo ($\alpha=1.5$) mass bias for $M=10^{14.6}, 10^{14.8},
    10^{15.0}$, and $10^{15.2}~\textup{M}_{\odot}$, respectively. The mass maps
    are reconstructed with $\lambda=2\,$. The darker grey area indicate a $5\%$
    bias and the lighter grey area indicate a $20\%$ relative mass bias. The
    error bar indicate the standard deviation of reconstructed mass with
    respect to the ellipticity $(a/c)$ over the range $[0.5,1]$.
}
\label{NoisyMassBiasCuspylambda2}
\end{figure*}
This is our result section on how \splinv{} performs reconstructing a noisy
shear field. We first analyzed how the $\lambda$ value affects the
reconstruction results in Sect. \ref{sec:oneHalo_noisy_lambda} and set a
fiducial value of $\lambda=2$. Then in Sect. \ref{sec:oneHalo_noisy_mass} and
Sect. \ref{sec:oneHalo_noisy_z} we show the result of mass and redshift
estimation from noisy reconstructions. In \ref{sec: Model bias}, we study the
potential model bias due to the difference between halo models in the universe
and those used in our dictionary.

\subsubsection{Performance with different $\lambda$}
\label{sec:oneHalo_noisy_lambda}

The performance of the our mass mapping algorithm may depend on the
regularization parameter $\lambda$ for the noisy case. We present the how
\splinv{} behaves differently with different $\lambda$ values in this section.

To determine the sparsity parameter $\lambda$ in equation~\eqref{eq:lasso} that
optimizes reconstruction results, we perform various single halo reconstruction
of shear field produced by an isotropic halo of mass $10^{14.6}, 10^{14.8}$,
and $10^{15.0}$ respectively, at an intermediate redshift ($z=0.2425$) in the
center of in a $48 \times 48$ pixelized grid covering $98 \,\text{arcmin}
\times 98\, \text{arcmin}$ of sky area. Examining the density plots (Figs.
\ref{ViolinPlot14.6}---\ref{ViolinPlot15.0}) for different values of $\lambda$,
we find that while ``the best'' $\lambda$ parameter potentially exists for each
case, a smaller $\lambda$ value tend to make \splinv{} to provide a smaller
mass estimation than those provided by \splinv{} with a larger $\lambda$, most
likely due to a smaller $\lambda$ relaxes the sparsity condition as can be seen
in equation ~\eqref{eq:lasso}.

We conclude that a relatively higher $\lambda$ should more strongly enforce the
sparsity condition, while effective making a cutoff for false detections with
small masses. For this same reason, larger $\lambda$ reduce the probability of
detecting halos with small masses.

We find that the optimal value of $\lambda$ depends on both the mass and the
redshift of the halo, as reconstruction of a halo with larger SNR (higher mass
and lower redshift) prefers a larger $\lambda$. From this, we conclude that we
should find an optimized $\lambda$ for interval targeted detection
mass/redshift, and then recursively apply \splinv{} to detect galaxy halo in
each mass/redshift interval. For detecting halos with relatively smaller masses
($~10^{14.6}~\textup{M}_\odot$), authors recommend using $\lambda = 2$ and for
halos with large masses $~10^{15.2}\textup{M}_\odot$, we recommend using
$\lambda=4$. We leave the study on the optimal setup of $\lambda$ using
realistic ray-tracing simulations \citep{raytracingTakahashi2017} to future
works. More specifically, what our findings can be concluded as:
\begin{enumerate}
    \item For simulations with small halo masses ($10^{14.6} M_\odot \leq M\leq
        10^{14.8}\,M_\odot$), we find a $\gtrapprox 18\%$ positive mass estimation bias
        exists for reconstruction with $\lambda>=2.5$. This is possibly due to
        a form of Eddington Bias (e.g., see \cite{Kelly:2007jy} and
        \cite{Eddington1913}), where only halo's shear signal boosted by noises
        gets detected which are then confused with halo with a large mass.
    \item For halos with larger mass ($M\geq 10^{15.0} M_{\odot}$) we do not
        find significant mass bias with $\lambda\geq 3$ due to the high SNR of
        these halos.
    \item For halos with larger mass, \splinv{} slightly underestimate halo
        masses with small $\lambda$ ($\lambda \leq 2.5$). This is possibly
        because the strong signal of higher mass halos may be cause the
        sparsity condition of \splinv{} to fail and be construed by our
        algorithm as caused by multiple halos.
\end{enumerate}

Because we found \splinv{} with $\lambda=2$ performs well with smaller mass
halos (which are more abundant in the universe) and only suffers overestimation
slightly, we set $\lambda =2$ as fiducial setup and put results with $\lambda =
4$ into Appendix~(\ref{lambda=4,append}).

\subsubsection{Mass Estimation}
\label{sec:oneHalo_noisy_mass}

\begin{figure*}
\centering
\includegraphics[width=0.9\textwidth]{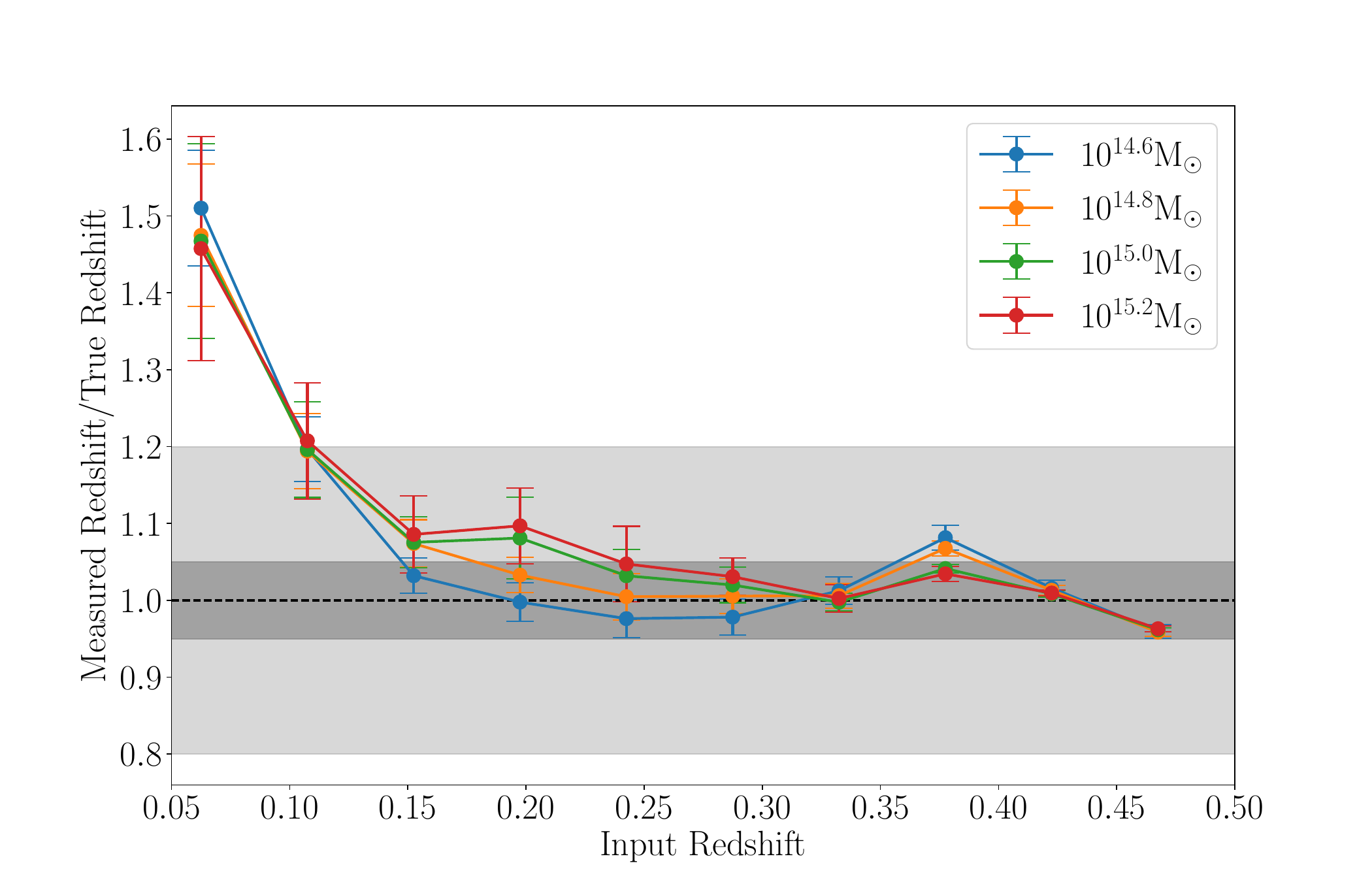}
\caption{
    NFW halo ($\alpha=1$) redshift estimation for $M=10^{14.6}$, $10^{14.8}$,
    $10^{15.0}$, and $10^{15.2}~\textup{M}_{\odot}$, respectively. The mass
    maps are reconstructed with $\lambda=2$. The darker grey area indicate a
    $5\%$ bias and the lighter grey area indicate a $20\%$ relative mass bias.
    The error bar indicate the standard deviation of reconstructed mass with
    respect to $\frac{a}{c}$ over the range $[0.5,1]$.
    }
    \label{NoisyRedshiftBiasNFWlambda2}
\end{figure*}

In this section, we present reconstruction results for triaxial NFW
($\alpha=1$) and cuspy NFW ($\alpha=1.5$) halos.

We simulate $100$ halos with $10$ different shapes (from $\frac{a}{c}=1$ to
$\frac{a}{c}=0.5$) and $10$ different redshifts (from $z=0.0625$ to
$z=0.4675$), and for each halo, we generate $500$ realizations of observational
noise as described in Section~\ref{sec:onHalo_noi}.

In Figs.~\ref{NoisyMassBiasNFWlambda2} and \ref{NoisyMassBiasCuspylambda2}, we
show the estimated relative mass biases for the two types of halo with ellipticity
($\frac{a}{c}$) ranging from $1$ to $0.5$ reconstructed with dictionary
generated numerically with the same mass and concentration parameter, but with
$\frac{a}{c} = 1$ (isotropic). At lower redshifts, corresponding to stronger
lensing signal, the choice of $\lambda=2$ gives halo mass estimations with
error less than $10\%$ or even better. Observing the first panel in
Figs.~\ref{NoisyMassBiasNFWlambda2} and \ref{NoisyMassBiasCuspylambda2}, we
again see the effect that, because the shear produced by the underlying halo
was too small, even with $\lambda = 2$, only the halos whose shear was boosted
by noise gets picked up by our algorithm, resulting in an overestimation.
However, the non-monotonous pattern in the second panel of
Figs.~\ref{NoisyMassBiasNFWlambda2} and \ref{NoisyMassBiasCuspylambda2}
indicates that one could potentially optimize the value of $\lambda$ for each
halo at each redshift. We also see that for reconstruction of more massive
halos results in an underestimation of masses. This could be caused by the fact
that a smaller $\lambda$ enforces a weaker sparsity condition and \splinv{} may
confuse the large signal due to the massive halo as signal generated by two
separate halos. Another trend we find is that, the smaller cuspy NFW halos are
generally more significantly affected by noise and hence will tend to have
larger mass estimation bias.

Additionally, we find that there is a small $\frac{a}{c}$ dependence on the
estimated mass bias. However, this dependence is not nearly as strong as that
in the noiseless case, which tells us that we should focus on optimizing the
value of $\lambda$ or other detection strategies before we try to include other
parameters (like the triaxiality of halo models or its rotation) that
complicate our dictionary space.

\subsubsection{Redshift Estimation}
\label{sec:oneHalo_noisy_z}

\begin{figure*}
\centering
\includegraphics[width=0.9\textwidth]{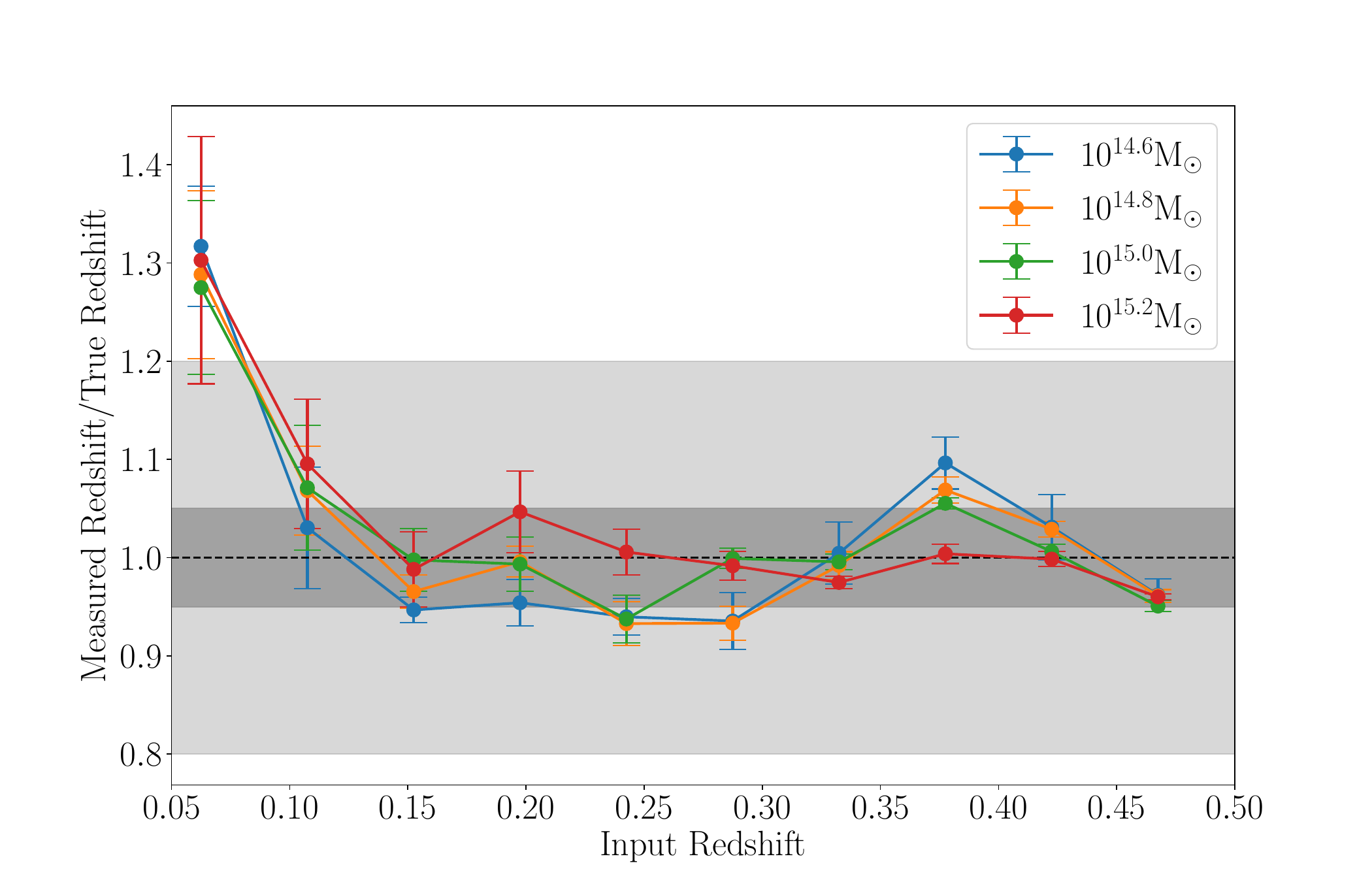}
\caption{
    Cuspy NFW halo ($\alpha=1.5$) redshift estimation for $M=10^{14.6},
    10^{14.8}, 10^{15.0},$ and $ 10^{15.2}$, respectively. The mass maps are
    reconstructed with $\lambda=2$\,. The darker grey area indicate a $5\%$
    bias and the lighter grey area indicate a $20\%$ relative mass bias. The
    error bar indicate the standard deviation of reconstructed mass with
    respect to $\frac{a}{c}$ over the range $[0.5,1]$.
    }
    \label{NoisyRedshiftBiasCuspylambda2}
\end{figure*}
Here we present the results of redshift estimations from noisy reconstructions
in Figs.~\ref{NoisyRedshiftBiasNFWlambda2} and
\ref{NoisyRedshiftBiasCuspylambda2}. We note that the slight overestimation of
redshift for halos with low redshifts in the figures is due to the discrete nature and
the lower boundary of the redshift bins: there cannot be an
underestimation for redshifts for these halos. Other than this, we observe that the
amplitude of the relative redshift estimation bias is consistently below $5\%$,
with no significant dependence on the shape ($\frac{a}{c}$ value) of the halo.

\subsubsection{Model Bias}
\label{sec: Model bias}
In the previous sections, we focus on the cases where halo model used for
reconstruction is the same as those used to create the shear field. In this
section, we study the potential model bias due to the systematic difference
between halo models in the universe and those used in our dictionary. Following
the previous sections, we are using isotropic models in our model dictionary.
In this section, we study the mass and redshift estimation under the condition
that the dictionary used for construction does not match the underlying halo in
the simulation that produces the shear field.

\begin{figure*}[!ht]
\centering
\includegraphics[width=1.0\textwidth]{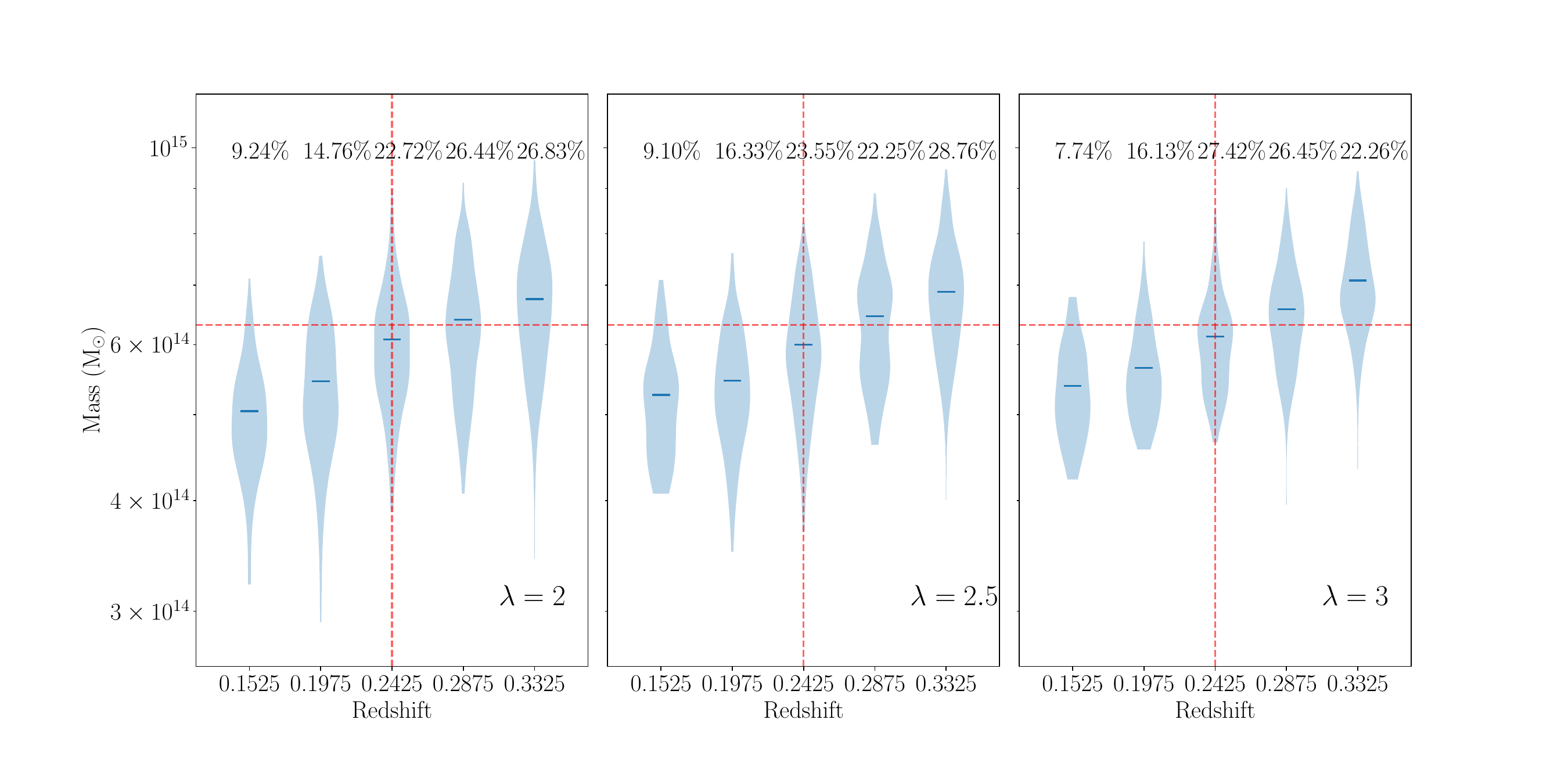}
\caption{
    Detected halo number densities of cuspy NFW halo with mass
    $10^{14.8}~\textup{M}_\odot$ using NFW halo as dictionaries. The blue
    shaded area indicate the number density of detected mass that correspond to
    the indicated mass and redshift. The percentage above one probability
    distribution represents the percent of total correct estimation
    corresponding to the respective redshift. The true detection probabilities
    are 78\%, 69\%, and 62\% for each value of $\lambda$ respectively.
    }
    \label{violin_cuspyfield_nfwdic}
\end{figure*}

In Fig.~\ref{violin_cuspyfield_nfwdic}, we show the effect of systematic error
due to the models used for mass map reconstruction being different. More
specifically, we show the result of estimating mass of a cuspy halo with the
assumption that the underlying mass field is composed of NFW halos. Comparing
Fig.~\ref{violin_cuspyfield_nfwdic} with Fig.~\ref{ViolinPlot14.8}, we find
that although we used the ``wrong'' dictionary in
Fig.~\ref{violin_cuspyfield_nfwdic}, the reconstructed result still resembles
that in Fig.~\ref{ViolinPlot14.8}.

\begin{figure*}[!ht]
\centering
\includegraphics[width=1.0\textwidth]{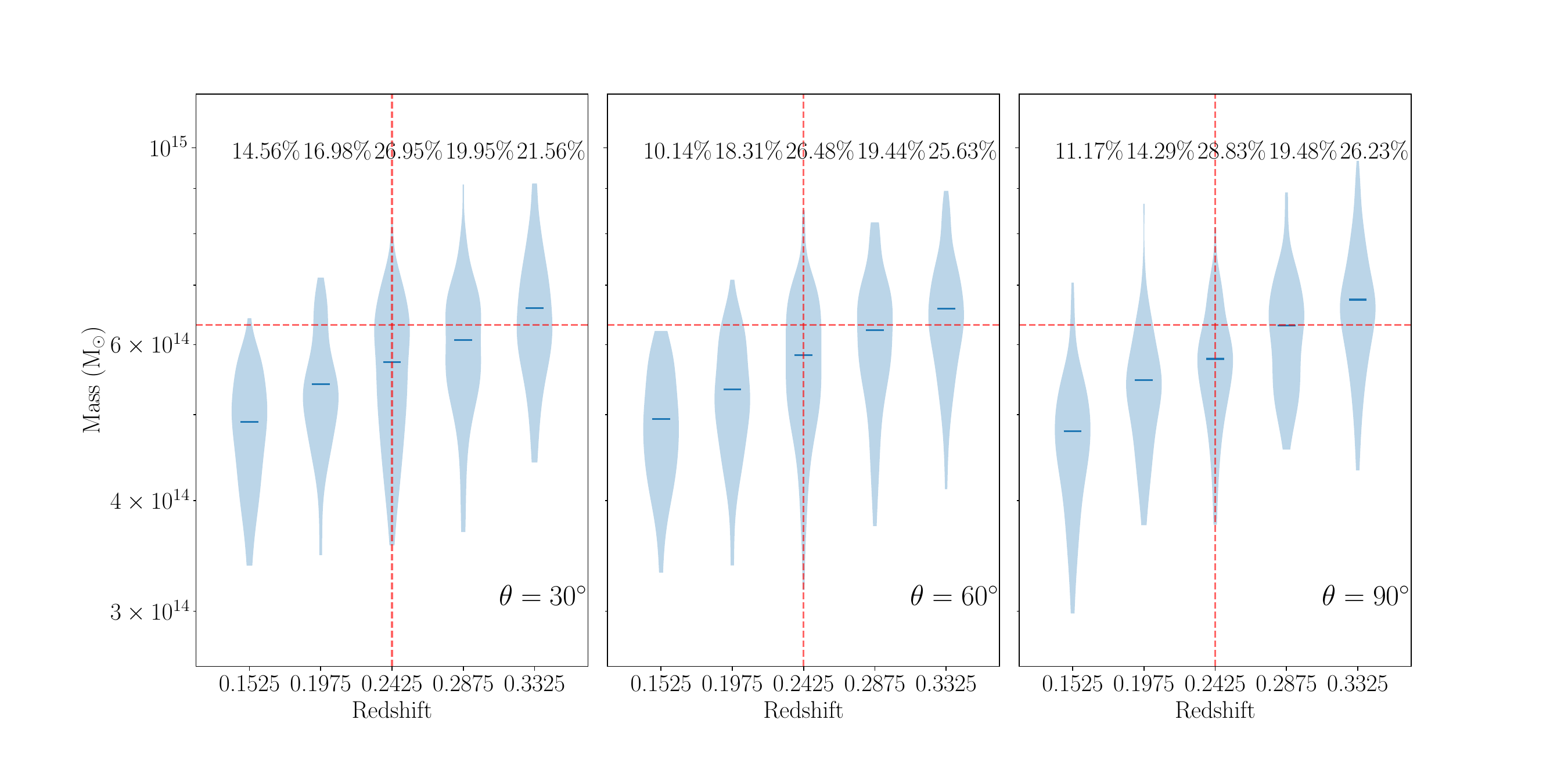}
\caption{
    Detection number density plot of NFW halo with $\frac{a}{c}=0.5$ with mass
    $10^{14.8}~\textup{M}_\odot$, but with $\theta = 30^\circ, 60^\circ,
    90^\circ$. The shaded blue area indicate the number density of detected
    mass that correspond to the indicated mass and redshift. The percentage
    above one probability distribution represents the percent of total correct
    estimation corresponding to the respective redshift. The detection
    probabilities are 74\%, 71\%, and 77\% for each value of $\lambda$
    respectively.
    }
    \label{violin_rotation}
\end{figure*}

\begin{figure*}[!ht]
\centering
\includegraphics[width=1.0\textwidth]{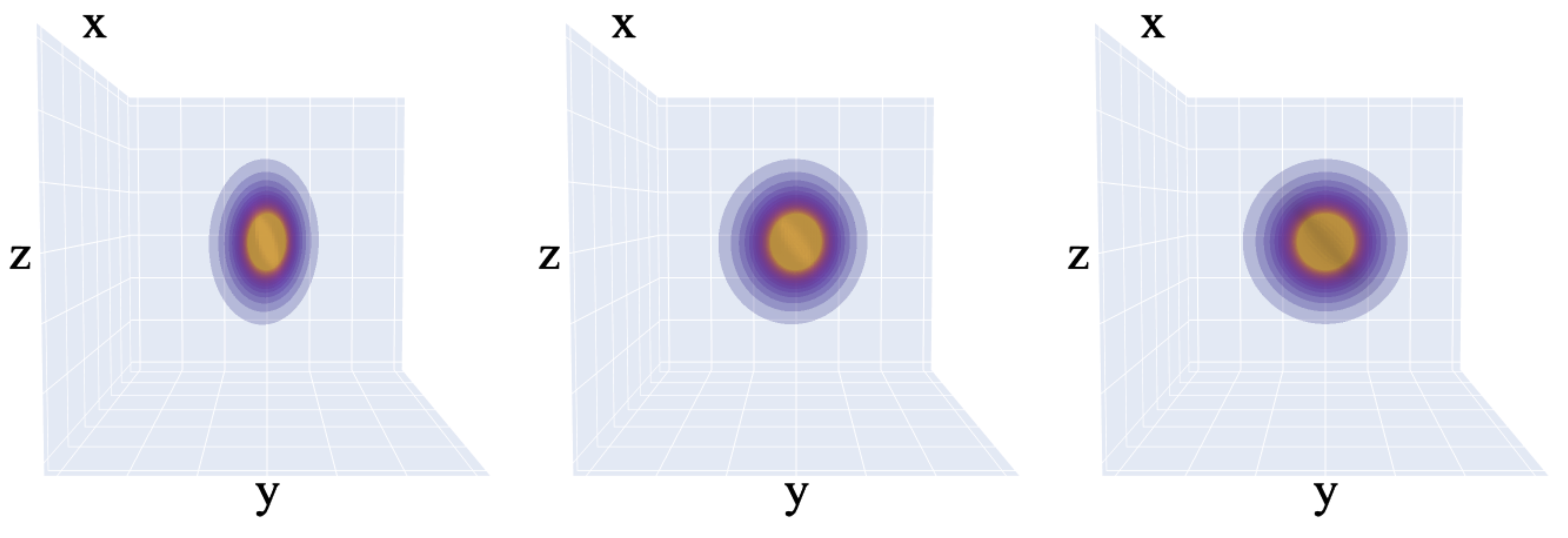}
\caption{
    A set of sample plot of halo density profile with halo with
    $\theta=30^\circ, 60^\circ,$ and $90^\circ$, respectively.
    \label{halo_rotation}
}
\end{figure*}

Next, we study whether the just using isotropic halo models in our dictionary
affect our ability to reconstruct highly anisotropic halos. Comparing
Fig.~\ref{violin_rotation} with the left panel of Fig.~\ref{ViolinPlot14.8}, we
see good agreement with reconstruction using isotropic halo model and we rotate
a highly anisotropic halo with $\frac{a}{c}=0.5$ in the polar direction for
$\theta = 30^\circ,60^\circ, 90^\circ$. A set of illustrative plots is shown in
Fig.~\ref{halo_rotation}. The results from this section and the previous one
indicate that, even when the true halo that constitutes the $\kappa$ map of the
universe may be anisotropic, one may still recover the underlying mass map
using isotropic models.

\section{MULTIPLE HALOS}
\label{sec:twohalos}

In this section we test our algorithm under the following conditions: (i)
Reconstruct mass map from noiseless shear field produced by 2 halos with
different separations; (ii) Reconstructing from noisy shear field produced by
multiple halos.

\subsection{Noiseless Two-halo Simulations}
\label{sec:noiseless 2 halo}
We start this series of simulation with two isotropic NFW halo of mass
$10^{14.8}~\textup{M}_\odot$ at the same redshift of $z=0.2425$ and change the
distance from $40$ arcmin to $0$ arcmin in a $48 \times 48$ pixelized grid
covering $98 \,\text{arcmin} \times 98\, \text{arcmin}$ of sky area.

Specifically, we decrease the distance between the two halos on the grid (as
measured by $ra$ and $dec$) by linear intervals, and perform the
reconstructions until the reconstruction fails, where either the number of halo
detected is wrong or the redshift estimate of either one of the halos is wrong
(meaning that the redshift estimation of the halos has to be \textit{exact}).
The other aspects of the simulation and reconstruction are identical to
\ref{sec:oneHalo_sim}.

An example of the reconstructed 3-D map is shown in
Fig.~\ref{fig:sample_2halo}. We observe that until the borderline of 4 arcmin,
the mass estimation of the two halos are consistently below $6\%$. Hence, we
should be concerned about significant mass bias due to halos closer or around
this cutoff in noisy reconstructions, where signals of two adjacent halo
combined with noise together creates a shear estimation that resembles a
different halo (false detection) which affects the mass and redshift estimation
of the original halo.

\subsubsection{Noisy Mutiple Halo Simulation}
\label{sec:noisy_multi_halo}

To test the performance of our algorithm in realistic multi-halo cases, we
consider the following simulation set-up. We use the same parameters adopted in
the previous sections but with a sky covering $256$~arcmin $\times256$~arcmin
area, corresponding to 128~pixels $\times$ 128~pixels. The center of the stamp
is set to $(\text{ra}, \text{dec})=(0^\circ, 0^\circ)$\,. The three cases are:
\begin{enumerate}
    \item The first halo has $M=10^{14.8}~M_\odot$ at $z=0.1975$ with
        $\text{ra}=4440''$ and $\text{dec}=5520''\,$, and the second halo has
        $M=10^{14.6}~{M}_\odot$ at $z=0.2875$ with $\text{ra}=-4440''$ and
        $\text{dec}=-4440''$\,. The distance are chosen to be far enough so
        that, even in the noisy simulations, there is little chance that two of
        the halos are falsely detected as one.
    \item The first halo has $M=10^{14.8}~\textup{M}_\odot$ at $z=0.2875$ with
        $\text{ra} = 4440''$ and $\text{dec} = 5520''$. The second halo has
        $M=10^{15.2}\textup{M}_\odot$ at $z= 0.3775$ with $\text{ra} = -4440''$
        and $\text{dec} = -4440''$. The third halo has $M=10^{14.6}~M_\odot$ at
        $z= 0.1975$ with $\text{ra} = 0''$ and $\text{dec} = 0''$\,. The
        distance are chosen to be far enough so that, even in the noisy
        simulations, there is little chance that two of the halos are falsely
        detected as one.
    \item Same with (ii) but we first perform a reconstruction with dictionary
        containing one halo with $M = 10^{15.2}~M_\odot$ (with a higher lambda)
        first, and then subtract the shear field produced by a realistic
        reconstruction result, containing information on estimated mass and
        redshift.
\end{enumerate}
The halo models used for the simulations and the reconstructions are both
isotropic in this section.

\begin{figure*}
\begin{center}
    \includegraphics[width=0.85\textwidth]{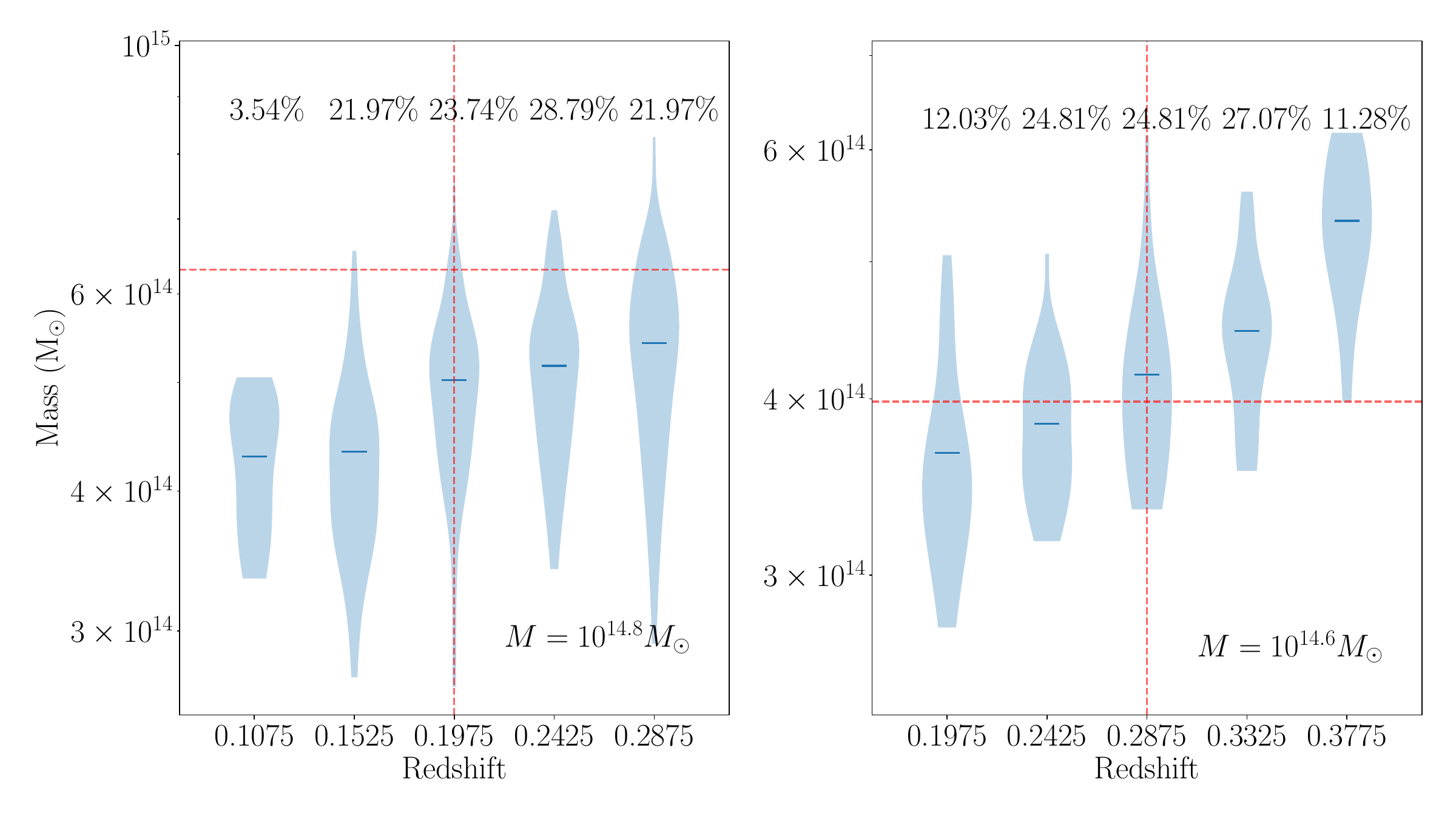}
\end{center}
\caption{
    Violin plot for reconstruction of halo with mass
    $M=10^{14.8}~M_\odot$ at $z=0.1975$ with ra$= 4440''$ and dec$=
    5520''$ and another halo which has mass $M=10^{14.6}M_\odot$ at
    $z=0.2875$ with ra$=-4440''$ and dec$=-4440''$\,. The red dashed curve
    represents the correct redshift and mass estimation. The percentage above
    one probability distribution represents the percent of total correct
    estimation corresponding to the respective redshift. The true detection
    probability was $79.2\%$ and $26.6\%$ respectively.
    }
    \label{Noisy2Halo_lbd2}
\end{figure*}

In Fig.~\ref{Noisy2Halo_lbd2}, we present result of case (i) with
dictionary composed of halos with $M = 10^{14.6}\textup{M}_\odot$ and with $M=
10^{14.8}$ with $\lambda=2$. While the estimation for the halo with $M =
10^{14.6}\textup{M}_\odot$ is accurate, we see an overestimation of halo mass
for the halo with $M = 10^{14.8}\textup{M}_\odot$. This is probably due to the
fact that we chose a generally small $\lambda$ for the halo with this mass and
a \splinv{} confuses the shear field produced by the
$10^{14.8}\textup{M}_\odot$ halo with that produced by the
$10^{14.6}\textup{M}_\odot$, but with a much larger mass to match the strength
of shear field. The slight decrease in detection probability is probability due
to increase in parameter space caused by one additional available choice of
atom which causes the gradient descent algorithm harder to converge.

\begin{figure*}
\begin{center}
    \includegraphics[width=0.85\textwidth]{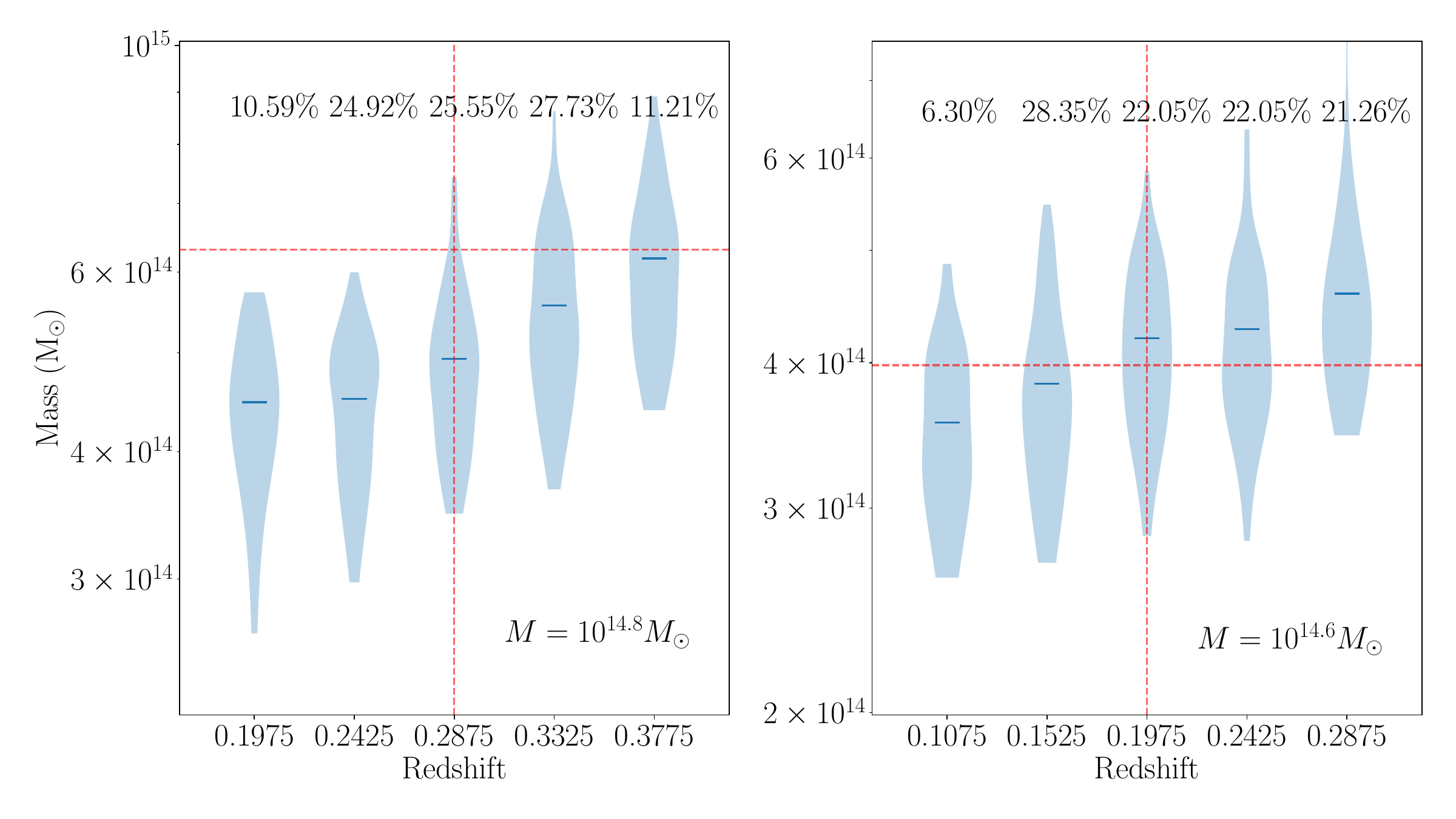}
\end{center}
\caption{
    Violin plot for reconstruction of halo with mass $M=10^{14.8}
    \textup{M}_\odot$ at $z=0.2875$ with $ra = 4440''$ and $dec = 5520''$,
    second halo which has mass $M=10^{14.6}\textup{M}_\odot$ at $z= 0.1925$
    with $ra = 0''$ and $dec = 0''$, and the last halo which has mass
    $M=10^{15.2}\textup{M}_\odot$ at $z= 0.3775$ with $ra = -4440''$ and $dec =
    -4440''$. We adopted dictionaries composing halos of mass $M =
    10^{14.6}\textup{M}_\odot$ and $M = 10^{14.8}\textup{M}_\odot$. The red
    dashed curve represents the correct redshift and mass estimation. The
    percentage above one probability distribution represents the percent of
    total correct estimation corresponding to the respective redshift. The true
    detection probability was $64.2\%$ and $50.8\%$ respectively.
    }
    \label{Noisy3Halo_lbd2_no_delete}
\end{figure*}

\begin{figure*}
\begin{center}
    \includegraphics[width=0.85\textwidth]{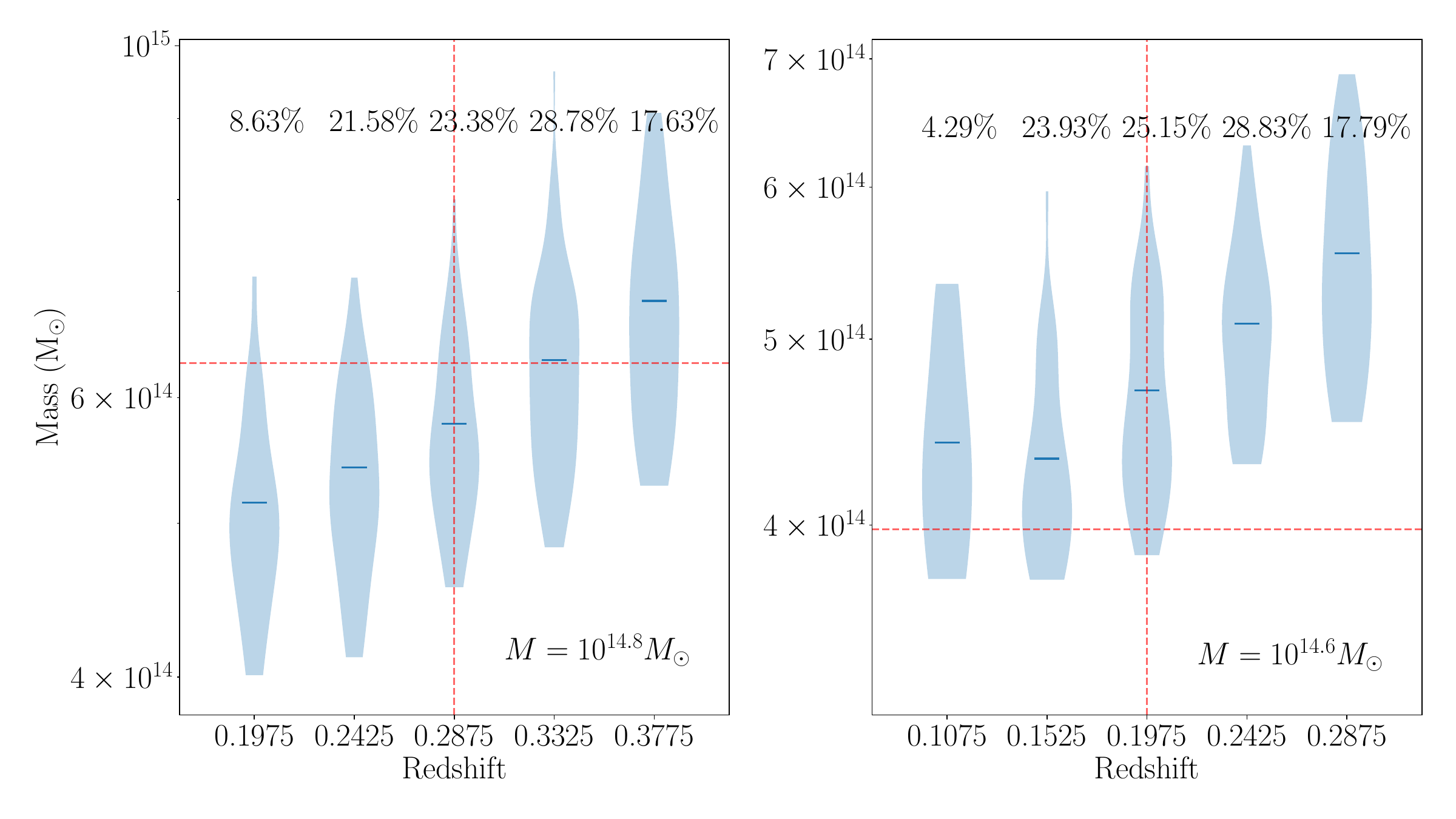}
\end{center}
\caption{
    Violin plot for reconstruction of halo with mass $M=10^{14.8}
    \textup{M}_\odot$ at $z=0.2875$ with $ra = 4440''$ and $dec = 5520''$,
    second halo which has mass $M=10^{14.6}\textup{M}_\odot$ at $z= 0.1925$
    with $ra = 0''$ and $dec = 0''$, and the last halo which has mass
    $M=10^{15.2}\textup{M}_\odot$ at $z= 0.3775$ with $ra = -4440''$ and $dec =
    -4440''$. We adopted dictionaries composing halos of mass $M =
    10^{14.6}\textup{M}_\odot$ and $M = 10^{14.8}\textup{M}_\odot$. We first
    perform a reconstruction with $\lambda = 4$ using dictionary with halo with
    $M = 10^{15.2}$ and then subtract one realistic reconstruction result. The
    red dashed curve represents the correct redshift and mass estimation. The
    percentage above one probability distribution represents the percent of
    total correct estimation corresponding to the respective redshift. The true
    detection probability was $62\%$ and $51.2\%$ respectively. }
\label{Noisy3Halo_lbd2_delete} \end{figure*}

In Fig.~\ref{Noisy3Halo_lbd2_no_delete} and Fig.~\ref{Noisy3Halo_lbd2_delete},
we show reconstruction results of the two smaller mass halo in case (ii) (a
good detection on the more massive halo can always be achieved with a high
value of $\lambda$). This is done in 2 ways: in the first method, we used
$\lambda=2$ and with dictionary composed of halos with $M =
10^{14.6}\textup{M}_\odot$ and with $M= 10^{14.8} ~M_\odot$ without modifying
the shear field. In the second method we first perform some detections of the
larger mass halo. Randomly select a set of mass and redshift estimation (we
used $M \approx 10^{15.11} M_\odot$ and $z=0.2875$ to produce the result), and
then proceed with reconstruction of the remaining two halos with $\lambda=2$
and with dictionary composed of halos with $M = 10^{14.6}\textup{M}_\odot$ and
with $M= 10^{14.8} ~M_\odot$.

We find that there is no significant difference in performance whether we
subtract the shear field produced by the large halo or not. However, the
underestimation in halo with $M=10^{14.8} ~M_\odot$ and overestimation in halo
with $M=10^{14.6}~M_\odot$ is still present. This result shows that, if we
focus on detecting smaller mass halos, we may safely use dictionaries of those
smaller mass halos without worrying about the shear field produced by large
mass halos to interfere with our detection, keeping in mind that any anomalous
large mass estimation may be caused by some large halo.

\section{SUMMARY}
\label{sec:Sum}
We preformed a set of systematic tests on the 3D mass map reconstruction
algorithm, \splinv{}, presented in \cite{massmap_Li2021}. \splinv{} can detect
NFW and cuspy NFW halos with $M = 10^{14.6}~M_\odot$ with less than $~5\%$ mass
bias in $0.0625\leq z \leq 0.2425$, $10^{14.8}~M_\odot$ with less than $~5\%$
mass bias in $0.0625\leq z \leq 0.4675$ and with less than $~20\%$ mass bias
for halo with $M= 10^{15.0} ~M_\odot$ and $M=10^{15.2}~M_\odot$ in the redshift
range $0.0625\leq z \leq 0.4675$. The redshift bias is consistently below $~5\%
z$ for the above halo masses in the range for $0.1525\leq z \leq 0.4675$. We
also demonstrated that rotations of triaxial halo models and systematic error
in halo modeling (e.g. we measure cuspy NFW halos with the assumption that
underlying mass filed of the universe is consisted of NFW halos) does not
affect reconstruction result significantly. Our multiple halo reconstruction
case demonstrated \splinv{}'s strong applicability to reconstruction to
observed shear catalogs measured by for example HSC and LSST in the future.

\section*{ACKNOWLEDGEMENTS}
This work is partially supported by Swarthmore College Honors Fellowship.

\section*{DATA AVAILABILITY}

The code used in this paper is available from
\url{https://github.com/mr-superonion/splinv/}.

\bibliographystyle{aasjournal}
\bibliography{citation}

\appendix

\section{Results with $\lambda=4$}\label{lambda=4,append}
\subsection{Mass Estimation}
\begin{figure*}
\begin{center}
    \includegraphics[width=0.9\textwidth]{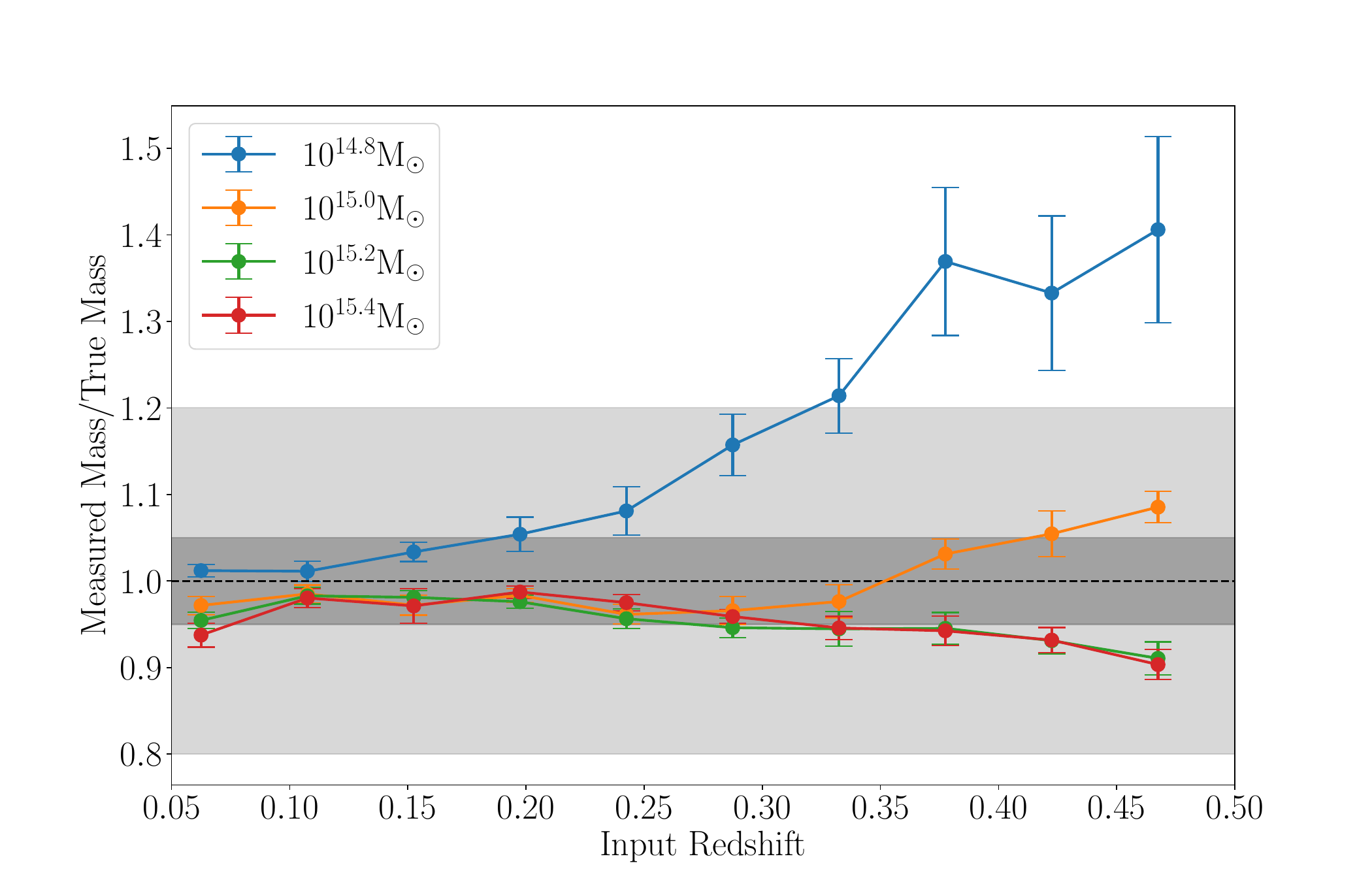}
     \caption{
     NFW halo Mass Bias for $M=10^{14.8}, 10^{15.0}, 10^{15.2},$ and $
     10^{15.4}$   $M_{\odot}$ respectively, reconstructed using $\lambda=4$.
     The darker grey area indicate a $5\%$ bias and the lighter grey area
     indicate a $20\%$ mass bias. The error bar indicate
    the standard deviation of reconstructed mass with respect to $\frac{a}{c}$ over the range
    $[0.5,1]$.
     }
     \label{NoisyMassBiasNFWlambda4}
\end{center}

\end{figure*}

\begin{figure*}
\begin{center}
\includegraphics[width=0.9\textwidth]{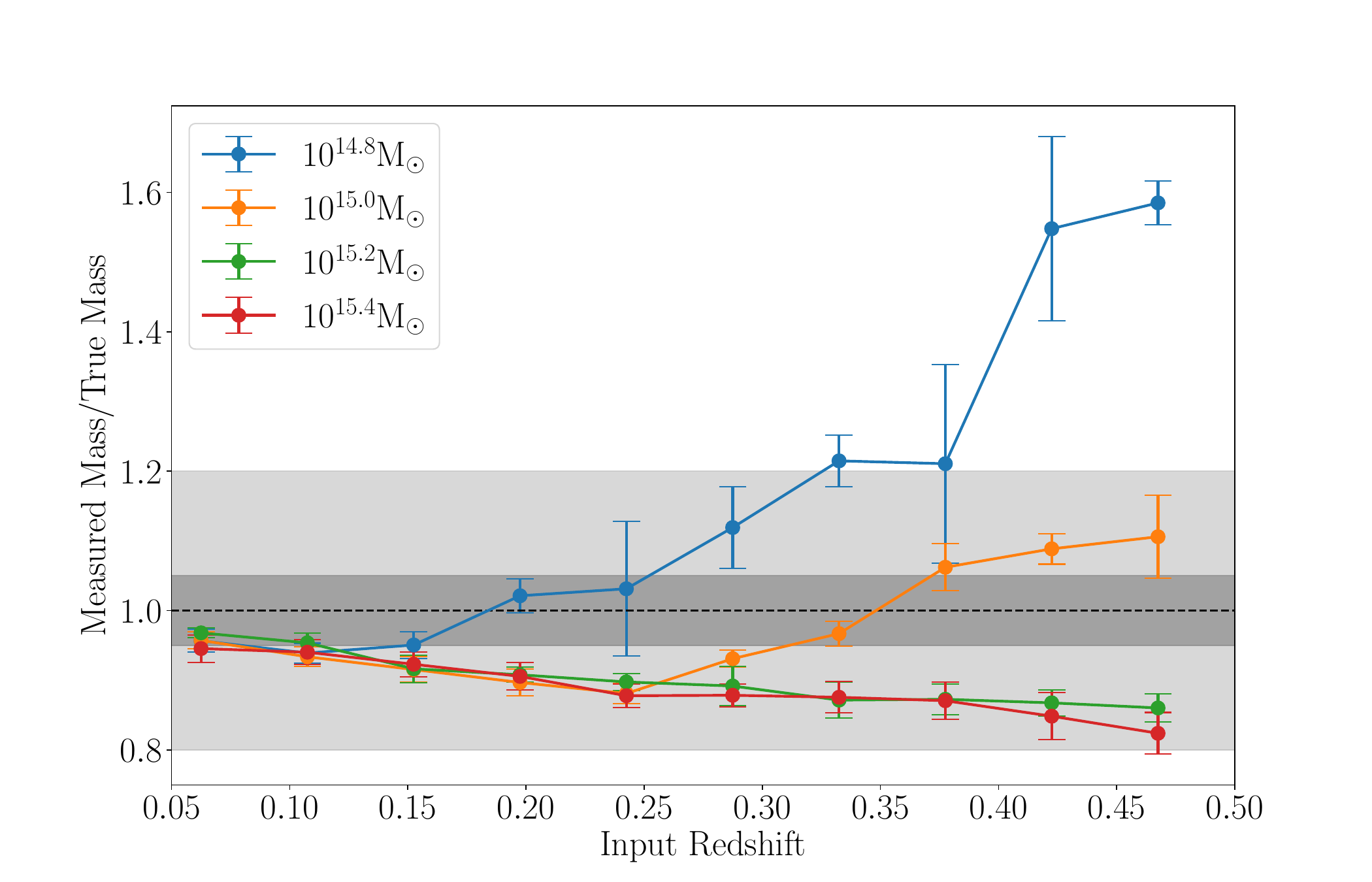}
\caption{
    Cuspy NFW halo Mass Bias for $M=10^{14.8}, 10^{15.0}, 10^{15.2},$ and $
    10^{15.4}$ $M_{\odot}$ respectively, reconstructed using $\lambda=4$. The
    darker grey area indicate a $5\%$ bias and the lighter grey area indicate a
    $20\%$ mass bias. The error bar indicate
    the standard deviation of reconstructed mass with respect to $\frac{a}{c}$ over the range
    $[0.5,1]$.
    }
    \label{NoisyMassBiasCuspylambda4}
\end{center}
\end{figure*}

Figs.~\ref{NoisyMassBiasNFWlambda4} and \ref{NoisyMassBiasCuspylambda4} show
the mass estimation of halos of masses $10^{14.8}, 10^{15.0}, 10^{15.2}$ and
$10^{15.4}$ $M_{\odot}$, reconstructed using $\lambda=4$. We observe for the $10^{14.8}M_\odot$ halos, while detection at lower redshift with a big $\lambda$ yields $\lessapprox 5\%$ mass
estimation bias, the
performance of \splinv{} decreases drastically as redshift of the halo
increases. With a larger $\lambda$, we see that the mass estimation for larger
mass improves, with performance of reconstructing NFW halos better than that of
cuspy NFW halos.

\subsection{Redshift Estimation}
For redshift estimations, we see a pretty similar result as in Sect. \ref{sec:oneHalo_noisy_z}, where the redshift estimation for halo of masses $10^{15.0}, 10^{15.2}$ and $10^{15.4}$ have consistently less than $5\%$ bias with $z>0.0625$. However, redshift estimation for halo with mass $10^{14.8} M_\odot$  for $z>0.3325$ shows above $40\%$ mass bias. This is probably due to the fact that, at this redshift level, halo with this mass are hard to detect with $\lambda=4$, causing we do not have enough data point to correct estimate the mass. 
\begin{figure*}
\begin{center}
\includegraphics[width=0.9\textwidth]{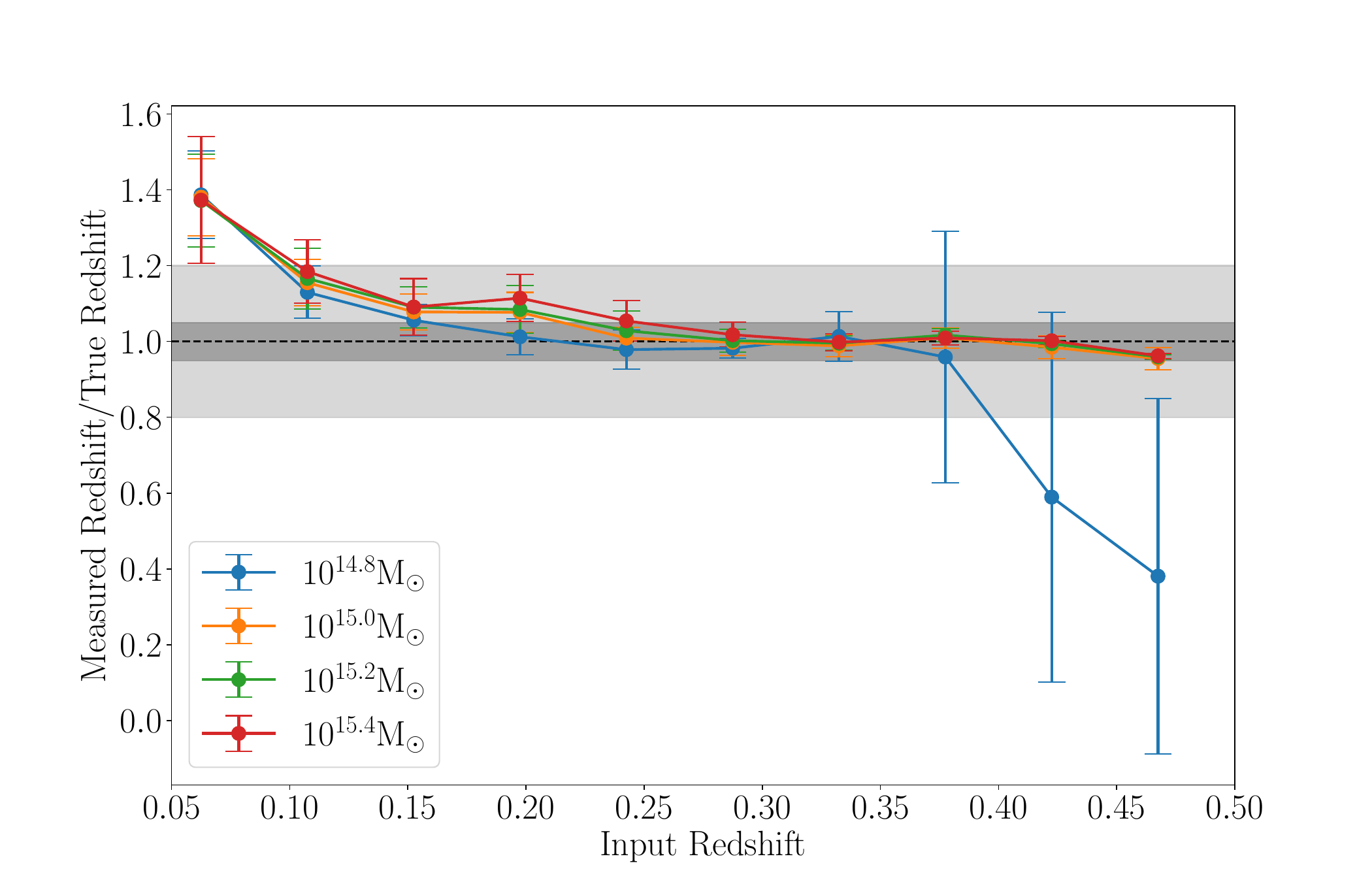}
\caption{
    NFW halo Redshift Estimation for $M= 10^{14.8}, 10^{15.0},
    10^{15.2}$ and $10^{15.4}$ $M_{\odot}$ respectively, reconstructed using
    $\lambda=2$. The darker grey area indicate a $5\%$ bias and the lighter
    grey area indicate a $20\%$ mass bias.
    }
    \label{NoisyRedshiftBiasNFWlambda4}
\end{center}
\end{figure*}

\begin{figure*}

\begin{center}
\includegraphics[width=0.9\textwidth]{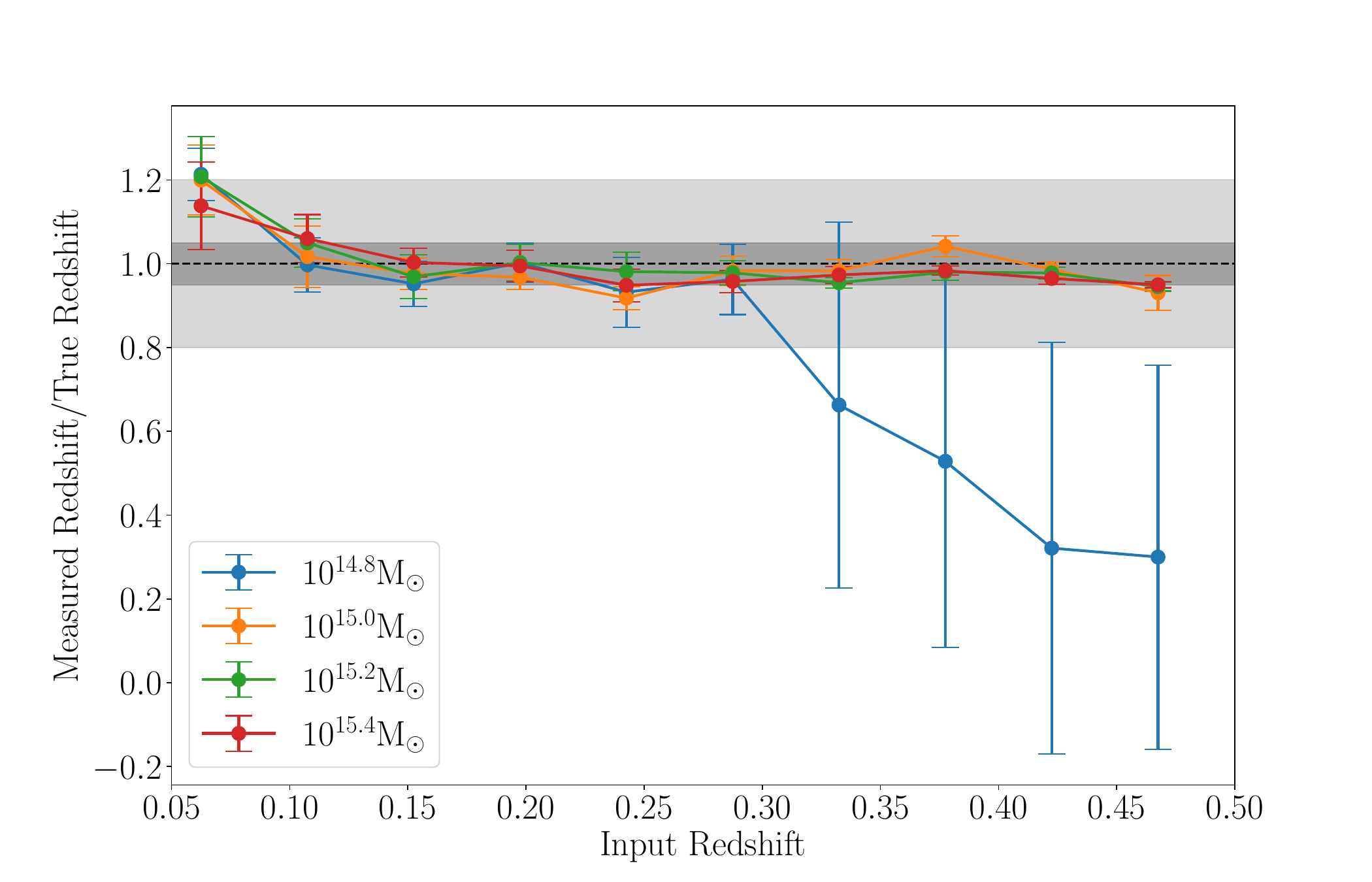}
\caption{
    Cuspy NFW halo Redshift Estimation for $M= 10^{14.8}, 10^{15.0},
    10^{15.2}$ and $10^{15.4}$ $M_{\odot}$ respectively, reconstructed using
    $\lambda=2$. The darker grey area indicate a $5\%$ bias and the lighter
    grey area indicate a $20\%$ mass bias.
    }
    \label{NoisyRedshiftBiasCuspylambda4}
\end{center}
\end{figure*}



\end{document}